\documentclass[a4paper,10pt]{article}

 \pdfoutput=1 

\usepackage[colorlinks=false, urlcolor=citecolor, linkcolor=citecolor, citecolor=citecolor]{hyperref}

\usepackage{amsmath, amssymb, amsthm, dsfont}
\usepackage[round]{natbib}
\usepackage{caption}
\usepackage{graphicx}

\usepackage[utf8]{inputenc}
\usepackage[OT1]{fontenc}
 \usepackage[english]{babel}

\usepackage{float}

\usepackage{algorithm,algpseudocode, algorithmicx}
\algnewcommand{\Inputs}[1]{%
  \State \textbf{inputs:}
  \Statex \hspace*{\algorithmicindent}\parbox[t]{.8\linewidth}{\raggedright #1}
}
\algnewcommand{\Initialize}[1]{%
  \State \textbf{initialize:}
  \Statex \hspace*{\algorithmicindent}\parbox[t]{.8\linewidth}{\raggedright #1}
}

\usepackage{titlesec}
\titleformat{\subsection}[runin]
       {\normalfont\bfseries}
       {\thesubsection}
       {0.5em}
       {}
       [.]
\titleformat{\section}
  {\normalfont\fontsize{12}{15}\bfseries}{\thesection}{1em}{}

\usepackage{geometry}
\title{\bf \Large Efficient inference for genetic association studies with multiple outcomes}
\author{H\'el\`ene Ruffieux$^{a,b, }$\thanks{To whom correspondence should be addressed (\href{mailto:helene.ruffieux@rd.nestle.com}{helene.ruffieux@rd.nestle.com}).}\;,
Anthony C. Davison$^b$,
J\"{o}rg Hager$^a$, Irina Irincheeva$^a$ \\
\normalsize $^a$Nestl\'e Institute of Health Sciences SA, Lausanne, Switzerland\\
\normalsize $^b$Ecole Polytechnique F\'ed\'erale de Lausanne (EPFL), Switzerland\\
}
\date{\normalsize \today}

\begin{document}

\maketitle

\begin{abstract}
{Combined inference for heterogeneous high-dimensional data is critical in modern biology, where clinical and various kinds of molecular data may be available from a single study. Classical genetic association studies regress a single clinical outcome on many genetic variants one by one, but there is an increasing demand for joint analysis of many molecular outcomes and genetic variants in order to unravel functional interactions. Unfortunately, most existing approaches to joint modelling are either too simplistic to be powerful or are impracticable for computational reasons. Inspired by \citet[][Bayesian Statistics 9]{richardson2010bayesian}, we consider a sparse multivariate regression model that allows simultaneous selection of predictors and associated responses. As Markov chain Monte Carlo (MCMC) inference on such models can be prohibitively slow when the number of genetic variants exceeds a few thousand, we propose a variational inference approach which produces posterior information very close to that of MCMC inference, at a much reduced computational cost. Extensive numerical experiments show that our approach outperforms popular variable selection methods and tailored Bayesian procedures, dealing within hours with problems involving hundreds of thousands of genetic variants and tens to hundreds of clinical or molecular outcomes.}\\

\end{abstract}

\section{Introduction}

Much current research in genetics focuses on combining heterogeneous data for the same samples. This is prompted by the increasing availability of diverse molecular data types from a single study and should lead to a more complete understanding of biological systems, as combined inference for data from different molecular layers might unravel regulatory interactions within and across layers \citep{civelek2014systems}. An example is \emph{protein quantitative trait locus} (pQTL) analyses to detect associations between hundreds of thousands of genetic variants and hundreds of proteomic expression levels. When associated with disease genetic variants, proteins are often regarded as intermediate phenotypes or molecular proxies for the disease of interest, as they may provide direct insights on biological processes underlying the clinical condition, and likewise for other molecules such as genes or metabolites involved in so-called eQTL or mQTL analyses.

An important goal of molecular QTL studies is to detect genetic variants with \emph{pleiotropic effects}, i.e., variants that regulate the expression levels of several molecules, as the genome location where they lie may initiate essential functional mechanisms \citep{breitling2008genetical}. As pleiotropy has been acknowledged as a central property of genetic variants causing phenotypic variation, efforts are being made to uncover its activity patterns and gain understanding of the shared biological processes it induces \citep{sivakumaran2011abundant, solovieff2013pleiotropy}. 
It is also of interest to identify those cases where the same molecule is simultaneously influenced by several genetic variants. This dual task requires a model that allows flexible selection, ideally performed jointly on the genetic variants and the molecular outcomes.

Although in practice univariate analyses still dominate, several proposals for joint modelling of multiple outcomes in genetic association problems have recently been made. \citet{flutre2013statistical} and \citet{zhou2014efficient} model the outcomes as multivariate responses having a matrix-variate distribution. The former rely on a Bayes factor framework to uncover the associations between a given genetic variant and any subset of outcomes, whereas the latter propose a linear mixed model whose random effect accounts for relatedness among individuals. While these methods show improvements over the fully marginal regression approach, they are restricted to problems with a few outcomes, because their models involve unstructured covariance matrices. \citet{o2012multiphen}'s MultiPhen method reverses the classical regression setup and fits a succession of models where each genetic variant is regressed on several outcomes. This eliminates the need to model large covariance matrices, but also entails a marginal treatment of the genetic variants. Moreover, MultiPhen does not penalize model complexity, which may cause instabilities when many outcomes are modelled. Molecular QTL problems are particularly complex, because in addition to the so-called $p \gg n$ paradigm, whereby the number of covariates (genetic variants) $p$ greatly exceeds the number of samples $n$, there is also a ``large $d$'' characteristic, as the number of responses (expression levels) $d$ is usually large. Methodologies to accommodate this are needed, especially when joint response modelling is sought. 

Two-stage procedures are natural approaches to association problems with both $p$ and $d$ large. 2HiGWAS \citep{jiang20152higwas} is essentially an implementation of the screening strategy of \citet{fan2008sure} in the context of longitudinal outcomes. It consists of a dimension reduction step, where each genetic variant is tested against each outcome, followed by a penalized regression recast into functional mapping in which only the genetic variants remaining after screening are involved. While the second stage of 2HiGWAS is an interesting approach to joint covariate and response modelling, the method is not tailored to typical molecular QTL analysis, as it is designed for outcomes measured over time. Also, the fully marginal first stage screening, if too stringent, may cause important predictors to drop out of the analysis and thus lead to false negatives. Instead of pruning the covariate set, \citet{wang2016block} summarize information at outcome level in the context of eQTL analyses. They precluster the expression levels using a block-mixture model and test for association between each genetic variant and the resulting clustering. This approach requires that the stability of the clustering and its functional relevance are carefully checked, as the group pattern chosen is critical to subsequent analysis. More generally, two-stage procedures often rely on ad-hoc thresholding decisions at the first stage which influence the conclusions of the second stage. 
 
The approaches sketched above use very diverse strategies to model predictor and outcome variables in high-dimensional settings. Trade-offs between realistic modelling and computational efficiency are inevitable, but two components seem critical to practical and powerful analyses: involving all variables in a single multivariate model, and maintaining interpretable inference. Unified selection of genetic variants and associated outcomes is then possible, unlike for most existing methods, whose focus is on selecting either predictors or outcomes. The Bayesian framework seems particularly suitable, as it offers flexible modelling possibilities in which biological beliefs may be naturally incorporated. 
The hierarchical regression approach of \citet{richardson2010bayesian}, HESS (hierarchical evolutionary stochastic search), is an appealing example of such approaches; it can identify associations between hundreds of covariates and up to a few thousand responses from a single model. \citet{jia2007mapping} and \citet{scott2012integrated} propose methods called BAYES and iBMQ (integrated Bayesian hierarchical model for eQTL mapping) based on models similar to that of HESS, but a major drawback of all three approaches is the lack of scalability of their Markov chain Monte Carlo inference procedures. Problems whose size corresponds to actual genome-wide association studies with molecular outcomes (with hundreds of thousands of genetic variants and hundreds to thousands of outcomes and individuals) are out of reach even for HESS, which is based on adaptive parallel tempering/evolutionary Monte Carlo techniques. We are unaware of any fully multivariate approach that can deal with such data within a reasonable time.

In this paper, we propose a Bayesian inference strategy that avoids sampling, via an algorithm that is fast and whose convergence is easy to monitor, while having performance comparable with Markov chain Monte Carlo approaches. We describe a variational inference procedure for a model similar to that of \citet{richardson2010bayesian}, comprising a series of parallel linear regressions, one for each response, combined in a hierarchical manner to leverage shared information. A spike-and-slab prior \citep{ishwaran2005spike} is used to induce sparsity of the regression coefficients, and the probability that a given covariate affects any response is modelled through a parameter that is shared across responses. Interpretable posterior quantities, such as the probability of association of each covariate-response pair, are produced. An efficient algorithm for this model is crucial, as its number of parameters can be very large. Our variational approach can update the parameters jointly in a tractable manner. \citet{carbonetto2012scalable} provide a good discussion of variational inference in the context of genetic association and propose a variational regression method called ``varbvs'', which can be seen as a single-response counterpart of our approach. In our multiple response setting, we show by simulation that our procedure is accurate and reliable, and is better than existing methods at selecting variables for very large problems. Hence, it offers clear added-value in practice: it enables complex and flexible Bayesian inference based on large batches of genetic data, without having to prune them beforehand. 

The paper is organized as follows. Section \ref{SecMod} describes the model, discusses its relation to earlier proposals, and presents a procedure to allow for sparsity control at both covariate and response levels. Section \ref{SecVB} gives an overview of variational Bayes approaches and describes our inference strategy. Section \ref{SecEQ} compares variational and Markov chain Monte Carlo inferences on the same model, also using direct approximations of posterior quantities. Section \ref{SecNS} describes numerical experiments for larger problems, comparing our method with several predictor selection methods, including the varbvs approach of  \citet{carbonetto2012scalable}, and with methods performing combined covariate and response selection, namely HESS \citep{richardson2010bayesian} and iBMQ \citep{scott2012integrated}. The section also presents a permutation-based comparison of our method with varbvs on a real mQTL problem. Section \ref{SecConcl} summarizes the discussion and highlights further possible developments. 

Although the applicability of our method is not restricted to any particular context, all the numerical experiments presented in this paper use settings tailored to genome-wide association studies. The data-generation schemes are designed to embody common biological assumptions and are described in Section \ref{SecPred}. The method is implemented in the publicly available R package \texttt{locus}.

\section{Model and earlier proposals}\label{SecMod}

Let $y=(y_1, \ldots, y_d)$ be a $n \times d$ matrix of $d$ centered responses and let $X$ be a $n \times  p$ matrix of $p$ covariates, for each of $n$ samples. The covariates are $p$ genetic variants, more precisely single nucleotide polymorphisms, SNPs, and the responses might represent $d$ gene, protein or metabolite expression levels for $n$ individuals, depending on whether an eQTL, pQTL or mQTL problem is considered. Our model is intended to accommodate all the constraints entailed by molecular QTL analyses; it is adapted from that of HESS \citep{richardson2010bayesian}. Suppose that 
\begin{eqnarray*}
y_t &\mid& \beta_t, \tau_t \sim \mathcal{N}_n\left(X\beta_t, \tau_t^{-1} I_n\right), \qquad\qquad\qquad \tau_t \overset{\text{ ind}}{\sim} \textup{Gamma}(\eta_t, \kappa_t)\,, \qquad\qquad\qquad t = 1, \ldots, d\,,\\
\beta_{st} &\mid& \gamma_{st}, \tau_t, \sigma^2 \sim \gamma_{st}\,\mathcal{N}\left(0, \sigma^2\,\tau_{t}^{-1}\right) + (1-\gamma_{st})\,\delta_0 \,,\qquad\qquad\qquad \sigma^{-2} \sim \textup{Gamma}(\lambda, \nu)\,,\quad\\
\gamma_{st}&\mid& \omega_{s} \overset{\text{ iid}}{\sim} \text{Bernoulli}\left(\omega_{s}\right), \qquad\qquad\qquad\qquad \omega_s  \overset{\text{ ind}}{\sim} \textup{Beta}(a_s, b_s)\,,\qquad\qquad\qquad\quad   s=1, \ldots, p\,,
\end{eqnarray*}
where $\delta_0$ is the Dirac distribution. Each response, $y_t$, is related linearly to the covariates and has a specific precision, $\tau_t$. The responses are conditionally independent across the regressions, but dependence among responses associated with the same covariates is captured through the prior specification of parameters $\omega_s$ and $\sigma^{-2}$, which are shared across the responses. This formulation circumvents modelling the covariance between the responses, which is infeasible when $d$ is large. Each covariate-response pair has its own regression parameter, $\beta_{st}$, for which sparsity is induced using a spike-and-slab prior. The binary parameter $\gamma_{st}$ acts as a ``pair selection'' indicator; covariate $X_s$ is associated with response $y_t$ if and only if $\gamma_{st}=1$. The parameter $\sigma$ represents the typical size of nonzero effects and is modulated by the residual variance, $\tau_t^{-1}$, of the response concerned by the effect. The parameters $\gamma_{s1},\ldots, \gamma_{sd}$ specify the response(s) associated with $X_s$ and are identically distributed as Bernoulli with common parameter $\omega_s$. Thus, $\omega_s$ controls the proportion of responses associated with covariate $X_s$. The goal of inference is variable selection. Selection of predictors can be performed by ranking the posterior means of $\{\omega_s\}$ and selection of covariate-response pairs can be performed by ranking the posterior probabilities of inclusion, i.e., the posterior means of~$\{\gamma_{st}\}$.

Our model differs from that of \citet{richardson2010bayesian} in two respects. One concerns the treatment of the regression coefficient parameters $\beta_{st}$: we use independent priors, whereas \citeauthor{richardson2010bayesian} rely on g-priors \citep{zellner1986assessing}. The main motivation for our choice is that the effects of genetic variants on a given outcome can be understood as causal, since no retroactive process can affect the variants, and they can take place at locations of the genome that are far apart, so their correlation structure need not reflect the spatial correlation of the SNPs; see \cite{guan2011bayesian}. 
\citet{jia2007mapping} also rely on independent priors for the regression coefficients of BAYES, but they model the latter with a mixture of two normal distributions rather than a spike-and-slab prior and impose a residual variance parameter that is common to all responses. This stringent assumption may represent a weakness of their proposal. 

The second difference concerns the third level of the model. \citet{richardson2010bayesian} opt for a quite complex specification, in which
\begin{equation}\label{EqPriorHess}\omega_{st}=\rho_s\omega_t\,, \qquad\quad \omega_t \sim \textup{Beta}(a_t, b_t)\,, \qquad\quad  \rho_s \sim \textup{Gamma}(c_s, d_s)\,,\qquad\quad 0 \leq \omega_{st} \leq 1\,.\end{equation}
In their case, the inclusion probability of $X_s$ for response $y_t$ is modelled through $\omega_t$; it is specific to that response but can be regulated using the parameter $\rho_s$, common to all responses. 
\citet{jia2007mapping} and \citet{scott2012integrated} propose other variants for this prior. The former choose a treatment similar to ours, with $\omega_{st} \equiv \omega_s \sim \text{Dirichlet}(1,1)$, and the latter consider an additional level of hierarchy, 
\begin{equation}\label{EqPriorIbqm}
\omega_{st} \mid  a_s, b_s, \pi_s \sim \pi_s \delta_0 + (1-\pi_s)\text{Beta}(a_s, b_s)\,, \qquad\qquad\quad \pi_s \sim \text{Beta}(a_0, b_0)\,,
\end{equation}
with $a_s \sim \text{Exp}(\lambda_a)$ and $b_s \sim \text{Exp}(\lambda_b)$. Our choice $\omega_{st}\equiv\omega_s \sim \textup{Beta}(a_s, b_s)$ is partly driven by our wish to design a simpler model and partly by practical considerations, since it ensures a closed form for our variational algorithm, unlike with (\ref{EqPriorHess}). While such a formulation was mentioned by \citet{richardson2010bayesian} and by \citet{scott2012integrated}, they did not pursue it because of concerns regarding its ability to control for multiplicity. Indeed, our model inherently enforces sparse associations as the number of responses, $d$, increases, but no control is achieved when the number of covariates, $p$, grows. We address this below by providing a procedure to induce a correction through the prior of $\omega_s$. 

\begin{figure}
\centering
  \noindent\includegraphics[scale=1]{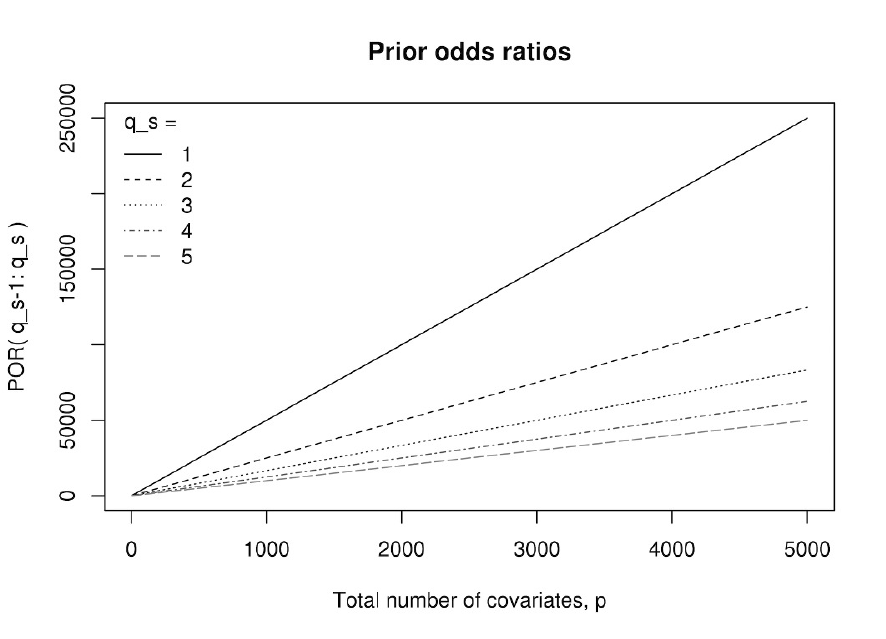} 
  \caption{\small Prior odds ratios, $\text{POR}(q_s-1:q_s)$, for $q_s=1,\ldots, 5$, $a_s$ and $b_s$ as in (\ref{EqHyperAdjust}), $d=100$, $p^*=2$, and for a total number of covariates ranging from $p=5$ to $5,000$; see \citet{scott2010bayes} for a similar visualization of prior odds ratios in a single response context.}\label{FigPOR}
\end{figure}

Part of the flexibility of our model comes from the fact that the hyperparameters, $a$, $b \in \mathbb{R}^p_+$ (for $\omega$'s Beta prior), $\lambda$, $\nu \in \mathbb{R}_+$ (for $\sigma^{-2}$'s Gamma prior) and $\eta$, $\kappa \in \mathbb{R}^d_+$ (for $\tau$'s Gamma prior) are readily interpreted. One option is to set them based on external information regarding the likelihood of given associations, if available. For instance, to favour associations with covariate $X_s$, one can set $a_s$ and $b_s$ so that the prior proportion of responses affected by $X_s$, $\text{E}(\omega_s)$, is large. The use of such assumptions may be very efficient, but it may also skew the inference towards existing knowledge. In the simulations presented in this paper, we assume that the regression and variance parameters are exchangeable, i.e., that all covariates, and responses have the same prior propensity to be involved in associations, by selecting a single value for all components of $a$, $b$, $\eta$ and $\kappa$. Without favouring any covariate or response, however, we can control signal sparsity at the level of covariates by specifying (possibly through cross-validation) a prior average number of covariates, $p^*$, expected to be included in the model. Setting 
 \begin{equation}\label{EqHyperAdjust}
 a_s \equiv 1\,, \qquad\quad b_s \equiv d(p-p^*)/p^*\,, \qquad\quad 0 < p^* < p\,,
 \end{equation} 
 the prior probability that $X_s$ is associated with at least one response is
$$p\left( \cup_{t=1}^d\{\gamma_{st}=1\} \right) = 1 - \frac{\prod_{j=1}^d (b_s + d -j )}{\prod_{j=1}^d ( a_s + b_s + d -j )} = \frac{p^*}{p}\,,$$
and simpler models are favoured as $p$ increases. To see this, one can consider the prior odds ratio representing the support for a model 
to have an additional response associated with~$X_s$, i.e.,
\begin{equation}\label{EqPOR}
\text{POR}(q_s-1:q_s)=\frac{p\left( \sum_{t=1}^d\gamma_{st}=q_s-1\right)}{p\left( \sum_{t=1}^d\gamma_{st}=q_s\right)}= \frac{b_s + d - q_s }{ a_s + q_s -1 }\,,\qquad\qquad q_s=1, \ldots, d\,.
\end{equation}
Clearly, penalties arise and increase with the total number of responses in the model, $d$. Figure~\ref{FigPOR} displays $(\ref{EqPOR})$ for $q_s=1, \ldots, 5$ as a function of $p$ and indicates that, when $a_s$ and $b_s$ are specified as in $(\ref{EqHyperAdjust})$, the penalties also increase with the total number of covariates, $p$, therefore naturally adjusting for multiplicity. Moreover, the penalties are not uniform when moving from one to two responses associated with $X_s$, or from four to five, for instance. 

The experiment reported in Table \ref{TabAdj} confirms that adjustment takes place in practice. It considers problems with $p_0=20$ ``active'' covariates, i.e., those associated with at least one response, and an increasing number of ``noise'' covariates and it compares the regime with $a_s$ and $b_s$ set according to (\ref{EqHyperAdjust}) to an ``uncorrected'' regime with $a_s\equiv1$, $b_s\equiv 2d-1$, so that the prior mean number of responses associated with $X_s$ is $0.5$, i.e., $\text{E}(\omega_s) \equiv (2d)^{-1}$. The number of false positives, based on a posterior probability of inclusion greater than $0.5$, grows linearly with $p$ when the uncorrected model is used but remains roughly constant close to zero with correction (\ref{EqHyperAdjust}), giving a clear multiplicity adjustment. Other experiments confirmed strong sparsity control; the reported findings on real data should therefore be plausible when (\ref{EqHyperAdjust}) is used. 

\begin{table}
\begin{center}
\captionof{table}{\small Multiplicity adjustment at covariate level. The mean numbers of false positives (FP) and true positives (TP) obtained with the uncorrected and corrected regimes are compared for $p_0=20$ active covariates and an increasing number of noise covariates, $p-p_0$. The total number of responses is $d=25$. $64$ replicates were performed; standard errors are in parentheses.\label{TabAdj}}
\footnotesize
\begin{tabular}{rccccc}
  \hline
 $p$ & $50$ & $250$ & $500$ & $1,000$ & $2,500$ \\ 
  \hline
 Mean \# of FP \\Uncorrected & $0.61$ ($0.66$) & $5.23$ ($2.24$) & $10.58$ ($3.16$) & $22.38$ ($4.48$) & $52.61$ ($8.39$) \\ 
  Corrected & $0.77$ ($0.81$) & $0.70$ ($0.99$) & $0.61$ ($0.77$) & $0.39$ ($0.58$) & $0.44$ ($0.59$)\\ 
  Mean \# of TP \\
  Uncorrected & $19.95$ ($0.21$) & $20.00$ ($0.00$) & $19.98$ ($0.12$) & $20.00$ ($0.00$) & $19.94$ ($0.24$) \\ 
  Corrected & $19.97$ ($0.18$)& $19.91$ ($0.34$) & $19.81$ ($0.43$) & $19.77$ ($0.56$) & $19.38$ ($0.86$)\\ 
   \hline
\end{tabular}
\end{center}
\end{table}

\section{Variational inference}\label{SecVB}

Section \ref{SecMod} described some differences between our model and those of \citet[][BAYES]{jia2007mapping}, \citet[][HESS]{richardson2010bayesian} and \citet[][iBMQ]{scott2010bayes}, but a more fundamental distinction concerns the inference procedure. The three earlier methods rely on Markov chain Monte Carlo (MCMC) techniques and require massive computing resources when the dimensionality of the problem is large. We instead employ a variational inference procedure, which is deterministic and hence can be much cheaper \citep{ormerod2010explaining}. 

Instead of sampling from the joint posterior probability $p(\theta \mid y)$ of the parameter vector of interest, $\theta$, 
variational approaches proceed by replacing it by a tractable analytical approximation, $q(\theta)$. We focus on so-called \emph{mean-field} variational formulations \citep{xing2002generalized, attias2000variational} to construct such a class, i.e., we assume that $q(\theta)$ factorizes over some partition of $\theta$, $\{\theta_j\}_{j=1,\ldots, J}$,
$$q(\theta) = \prod_{j=1}^J q_j(\theta_j)\,;$$
no further assumption is made about the distribution, and in particular no constraint is imposed on the functional forms of the $q_j(\theta_j)$. Here, we consider the factorization 
\begin{eqnarray}\label{EqFact} 
q\left(\beta, \gamma, \tau, \sigma^{-2}, \omega\right) =  \left\{ \prod_{s=1}^p\prod_{t=1}^d  q(\beta_{st}, \gamma_{st}) \right\}\left\{\prod_{s=1}^p q(\omega_s)\right\}\left\{\prod_{t=1}^{d} q(\tau_t) \right\} q\left(\sigma^{-2}\right)\,,
\end{eqnarray}
and turn the inference into an optimization problem where $q(\theta)$ is obtained by minimizing its Kullback--Leibler divergence $\text{KL}\left(q \middle\| p\right)$ from the target distribution, $p(\theta \mid y)$. 
Because the marginal log-likelihood may be written as 
\begin{equation}\label{EqKL}
\log p(y) = \mathcal{L}(q) + \text{KL}\left(q \middle\| p\right)\,,
\end{equation}
where
$$\mathcal{L}(q) = \int q(\theta) \log\left\{\frac{p(y, \theta)}{q(\theta)}\right\}\text{d}\theta\,,\qquad \text{KL}\left(q \middle\| p\right) = - \int q(\theta) \log\left\{ \frac{p(\theta\mid y)}{q(\theta)}\right\} \text{d}\theta\,, $$
minimizing the Kullback--Leibler divergence amounts to maximizing $\mathcal{L}(q)$, which represents a lower bound for $\log p(y)$. To this end, we observe that
\begin{eqnarray}
\mathcal{L}(q) &=& \int \prod_{k=1}^J q_k(\theta_k) \left\{ \log p(y,\theta) - \sum_{k=1}^J\log q_k(\theta_k) \right\} \text{d}\theta_1\cdots\text{d}\theta_J\nonumber\\
&=& \int q_j(\theta_j) \left\{ \int \log p(y,\theta) \prod_{k\neq j} q_k(\theta_k) \text{d}\theta_k - \log q_j(\theta_j) \right\} \text{d}\theta_j + \text{cst} \nonumber\\
&=& \int q_j(\theta_j) \log\left\{\frac{p_{-j}(\theta_j; y)}{q_j(\theta_j)}\right\} \text{d}\theta_j + \text{cst}\,, \qquad\qquad\qquad\qquad\qquad\quad j =1, \ldots, J\,,\label{EqExpKL}
\end{eqnarray}
where $\text{cst}$ is constant with respect to $\theta_j$ and where we introduced the distribution
$$p_{-j}(\theta_j; y) = \text{cst} \times \exp\left[ \text{E}_{-j}\left\{ \log p(y, \theta) \right\}\right]\,,$$
with $\text{E}_{-j}\{\cdot\}$ denoting the expectation with respect to the distributions $q_k$ over all variables $\theta_k$, $k\neq j$. The right-hand side of (\ref{EqExpKL}) corresponds to the negative Kullback--Leibler divergence between $q_j(\theta_j)$ and $p_{-j}(\theta_j; y)$, plus a constant. Hence, assuming that the $q_k(\theta_k)$, $k\neq j$, are fixed, the distribution $q_j(\theta_j)$ which maximizes $\mathcal{L}(q)$ is $q_j(\theta_j) = p_{-j}(\theta_j; y)$, i.e., the maximum of $\mathcal{L}(q)$ occurs when
\begin{equation}\label{EqQj}
\log q_j(\theta_j) = \text{E}_{-j} \{\log p(y, \theta)\} + \text{cst}\,,\qquad\qquad j=1, \ldots, J\,.
\end{equation} 
The relations $(\ref{EqQj})$ give rise to cyclic dependencies among the densities $q_j(\theta_j)$. This suggests an iterative algorithm whose convergence can easily be monitored by evaluating changes in the lower bound $\mathcal{L}(q)$. Our choice (\ref{EqFact}) ensures that the coordinate updates can be derived in closed form; in particular, the semi-conjugacy of our model implies that the prior densities of all parameters are preserved by the variational densities. For instance, a spike-and-slab distribution with modified parameters is recovered at posterior level, $q(\beta_{st}, \gamma_{st}) = q(\beta_{st} \mid \gamma_{st}) q(\gamma_{st})\,,$ with 
$$\beta_{st} \mid \gamma_{st}=1, y\; \sim \mathcal{N}\left(\mu_{\beta, st}, \sigma^2_{\beta, st}\right)\,,\qquad \beta_{st} \mid \gamma_{st}=0, y\; \sim \delta_0\,,\qquad \gamma_{st} \mid y \sim \text{Bernoulli}\left(\gamma^{(1)}_{st}\right)\,,$$
where the \emph{variational parameters} $\mu_{\beta, st}$, $\sigma^2_{\beta, st}$, $\gamma^{(1)}_{st}$ are to be updated iteratively. Convergence is ensured by the convexity of $\mathcal{L}(q)$ in each of the $q_j(\theta_j)$ \citep[\S\S\,3.1.5, 3.2.4, 3.2.5]{boyd2004convex}. The algorithm and its derivation are given in Appendix \ref{AppB}.

\section{Empirical quality assessment of the variational approximation}\label{SecEQ}

\subsection{Tightness of the marginal log-likelihood lower bound}\label{SecLB}

In this section we evaluate the closeness of the variational density $q$ to the target posterior distribution by approximating the Kullback--Leibler divergence $\text{KL}\left(q \middle\| p\right)$. Because of relation $(\ref{EqKL})$, this amounts to assessing the tightness of the variational lower bound for the marginal log-likelihood, $\mathcal{L}(q)$. 
For small problems, the likelihood $p(y)$ may be accurately approximated using simple Monte Carlo sums. 
We have 
$$
p(y) = \int\cdots\int \mathrm{d}\omega\,\mathrm{d}\sigma^{-2}  \, \left\{ \prod_{s=1}^p\,p( \omega_s)\right\} \,p\left(\sigma^{-2}\right)\prod_{t=1}^d \left\{\sum_{\gamma_t \in \{0,1\}^p}  p\left(y_t\mid\gamma_t, \sigma^{-2}\right)  \prod_{s=1}^p\,p\left(\gamma_{st} \mid\omega_s\right) \right\}\,,
$$
with 
{
$$p\left(y_t\mid\gamma_t, \sigma^{-2}\right) = \begin{cases} \displaystyle (2\pi)^{-n/2}\, \Gamma\left(\frac{n}{2} + \eta_t\right) \frac{\kappa_t^{\eta_t}}{\Gamma(\eta_t)} \left(\kappa_t + \frac{\Vert y_t \Vert^2}{2}\right)^{-n/2-\eta_t}, \hspace{2cm} q_{\gamma_t}=0\,,
\\\text{}\\\displaystyle(2\pi)^{-n/2} \begin{vmatrix} V_{\gamma_t,\sigma^{-2}}\end{vmatrix}^{-1/2} \Gamma\left(\frac{n}{2} + \eta_t\right) \frac{\kappa_t^{\eta_t}}{\Gamma(\eta_t)} \left(\kappa_t +\frac{S^2_{\gamma_t}}{2} \right)^{-n/2-\eta_t} \left(\sigma^{-2}\right)^{q_{\gamma_t}/2}, \\ \hspace{10.1cm}\text{otherwise}\,,
\end{cases}$$
}
where $$q_{\gamma_t} = \sum_{s=1}^p \gamma_{st}\,,\qquad\quad V_{\gamma_t,\sigma^{-2}} = X_{\gamma_t}^TX_{\gamma_t} + \sigma^{-2}I_{q_{\gamma_t}}\,, \qquad\quad
S^2_{\gamma_t,\sigma^{-2}} = \Vert y_t\Vert^2 - y_t^T X_{\gamma_{t}} V_{\gamma_t,\sigma^{-2}}^{-1} X_{\gamma_{t}}^T y_t\,;$$
see Appendix \ref{AppC} for details. As no closed form is available for the remaining integrals, we use 
$$
p(y) \approx \frac{1}{I} \sum_{i=1}^I \prod_{t=1}^d \left\{ \sum_{\gamma_t \in \{0,1\}^p}  p\left(y_t\mid\gamma_t, \left(\sigma^{-2}\right)^{(i)}\right)  \prod_{s=1}^p\,p\left(\gamma_{st} \mid\omega_s^{(i)}\right)\right\}\,,
$$
where we independently generate
\begin{equation}\label{EqSMC}\left(\sigma^{-2}\right)^{(i)} \sim \text{Gamma}(\lambda, \nu)\,, \qquad \omega_s^{(i)} \sim \text{Beta}(a_s, b_s)\,,\qquad s=1, \ldots, p\,,\qquad i=1, \ldots, I\,.\end{equation}

\begin{figure}
\centering
  \noindent\includegraphics[scale=1]{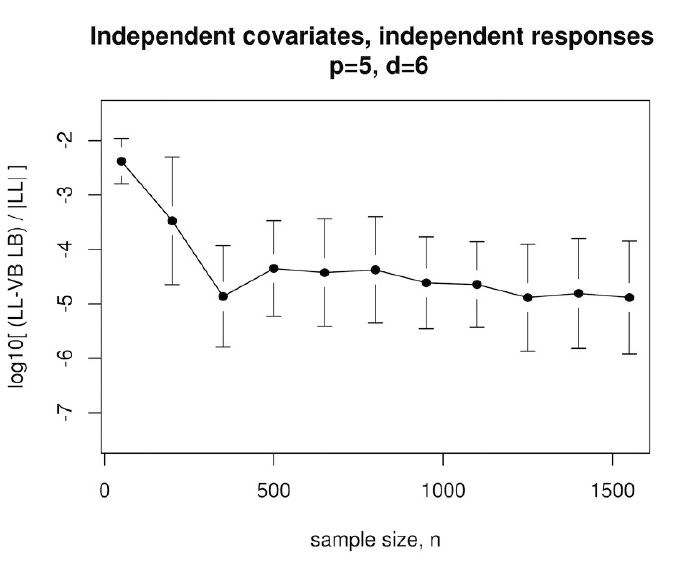}\hspace{0.4cm}
  \includegraphics[scale=1]{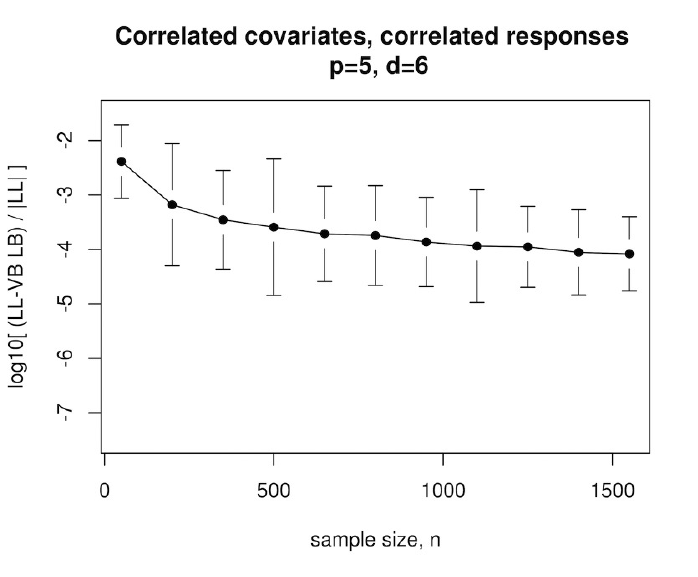}
    \caption{\small $\text{Log}_{10}$ relative difference between the marginal log-likelihood and the variational lower bound. Left: independent covariates and responses. Right: correlated covariates and responses, $\rho=0.75$. Problems with $p=5$ covariates, of which $p_0=3$ randomly selected as ``active'' (associated with at least one response), and $d=6$ responses, of which $d_0=3$ ``active'' (associated with at least one covariate). Each active covariate is associated with an additional active response with probability $0.25$ and explains on average $3.5\%$ of the variance of its corresponding response(s). The number of draws for the simple Monte Carlo approximations is $I=50,000$; the number of replicates for each sample size is $150$.}\label{FigLB}
\end{figure}

Figure \ref{FigLB} displays the relative difference $\left\{\log p(y) - \mathcal{L}(q)\right\}/\log p(y)$ for problems with $p=5$ covariates, $d=6$ responses and increasing sample sizes, $n$. In the left panel, the covariates are independent of each other, and so are the responses. In the right panel, the covariates are equicorrelated with correlation coefficient $\rho=0.75$, and so are the responses. In both cases, the mean relative difference is below $1\%$ with $n=50$ and seems to decrease as $n$ grows. Although we are not aware of any such study with which to benchmark our results, these values seem very small, suggesting that our variational distribution $q$ adequately reflects the target distribution $p$, at least for small problems. Likewise, the variational lower bound $\mathcal{L}(q)$ may be used as a proxy for the marginal log-likelihood when performing model selection; this use will be illustrated in Section \ref{SecReal}. The fact that the variational lower bound remains tight in the correlated data case is reassuring, as it suggests that the independence assumptions underlying the mean-field factorization of $q$ may only weakly impact the quality of the approximation.

\subsection{Comparison with Markov Chain Monte Carlo}\label{SecComp}

\begin{table}
\begin{center}
\captionof{table}{\small Variational Bayes (VB), MCMC and simple Monte Carlo estimates for $\beta$ and $\omega$ (components corresponding to noise averaged). Standard errors are in parentheses. \label{TabBetaOmega}}
\footnotesize
\begin{tabular}{rccccccc}
  \hline
  & & & Active & & & Inactive\\
 $10 \times$ & $\beta_{1,2}$ & $\beta_{2,1}$ & $\beta_{3,2}$ & $\beta_{4,1}$ & $\beta_{4,2}$ & $\beta_{\text{rest}}$ (avg) \\ 
  \hline
Truth & $-1.75$ & $2.87$ & $2.37$ & $3.73$ & $-4.76$ & $0.00$ \\ 
  VB & $-1.74$ ($0.01$) & $1.86$ ($0.01$) & $1.70$ ($0.02$) & $2.26$ ($0.01$) & $-3.48$ ($0.01$) & $0.02$ ($0.04$) \\ 
  MCMC & $-1.74$ ($0.33$) & $1.86$ ($0.32$) & $1.69$ ($0.34$) & $2.26$ ($0.32$) & $-3.48$ ($0.34$) & $0.02$ ($0.14$) \\ 
   \hline
      \hline
    & & &  Active & & & Inactive\\
 & $\omega_1$ & $\omega_2$ & $\omega_3$ & $\omega_4$ & &$\omega_{\text{rest}}$ (avg) \\ 
  \hline
  True prop. of active resp. & $0.2$ & $0.2$ & $0.2$ & $0.4$ & & $0$ \\
VB & $0.21$ ($0.14$) & $0.25$ ($0.15$) & $0.19$ ($0.14$) & $0.33$ ($0.17$) & & $0.03$ ($0.06$) \\ 
  MCMC & $0.23$ ($0.18$) & $0.26$ ($0.18$) & $0.21$ ($0.16$) & $0.35$ ($0.18$) & & $0.04$ ($0.08$) \\ 
  Simple Monte Carlo & $0.25$ & $0.26$ & $0.21$ & $0.35$ & & $0.05$ \\ 
   \hline
\end{tabular}
\end{center}
\end{table}

We complement our quality assessment by comparing several variational posterior quantities with those for MCMC inference on problems of moderate size. A fair comparison is not straightforward, as these two types of inference rely on stopping rules and convergence diagnostics of very different natures. While the convergence criterion for variational inference comes down to a tolerance to be prescribed, the ability of MCMC sampling to adequately explore the model space for a given chain length can be difficult to evaluate, and usually varies greatly with the problem size. To alleviate the risk of inaccurate MCMC inference, we run $10^5$ iterations and discard the first half. We also support our comparison with selected quantities approximated by simple Monte Carlo sums, namely, the posterior probability of inclusion of a covariate $X_s$ for a response~$y_t$,
\begin{eqnarray}
p( \gamma_{st}=1 \mid y ) &=& \frac{1}{p(y)} \;\frac{1}{I}\sum_{i=1}^I \left[ \prod_{t'\neq t} \left\{\sum_{\gamma_{t'} \in \{0,1\}^p}  p\left(y_{t'}\mid\gamma_{t'}, \left(\sigma^{-2}\right)^{(i)}\right)  \prod_{s'=1}^p\,p\left(\gamma_{s't'} \mid\omega_{s'}^{(i)}\right) \right\} \right.
\nonumber\\
&&\hspace{0.15cm}\times\left. \left\{\sum_{\gamma_{t} \in \{0,1\}^p:\; \gamma_{st}=1}  p\left(y_{t}\mid\gamma_{t}, \left(\sigma^{-2}\right)^{(i)}\right)  \prod_{s'=1}^p\,p\left(\gamma_{s't} \mid\omega_{s'}^{(i)}\right) \right\}\right]\,,\nonumber
\end{eqnarray}
and the posterior mean of $\omega_s$, controlling the proportion of responses associated with covariate~$X_s$, 
\begin{eqnarray}
\text{E}( \omega_{s} \mid y ) &=&  \frac{1}{p(y)} \;\frac{1}{I}\sum_{i=1}^I \;\omega_{s}^{(i)}\;\prod_{t=1}^d \left\{\sum_{\gamma_t \in \{0,1\}^p}  p\left(y_t\mid\gamma_t, \left(\sigma^{-2}\right)^{(i)}\right)  \prod_{s'=1}^p\,p\left(\gamma_{s't} \mid\omega_{s'}^{(i)}\right) \right\}\,,\nonumber
\end{eqnarray}
 with the samples $\left(\sigma^{-2}\right)^{(i)}$ and $\{\omega_s^{(i)}\}$ generated as in $(\ref{EqSMC})$, with $I=2 \times 10^5$ draws.
 
Table $\ref{TabBetaOmega}$ reports the variational, MCMC and simple Monte Carlo estimates of $\beta$ and $\omega$ for a problem with $p=8$ covariates, $d=5$ responses for $n=250$ samples, and with each nonzero association explaining on average $13.5\%$ of response variance. 
The estimates all agree closely. Those of the five active regression coefficients, $\beta_{1,2}$, $\beta_{2,1}$, $\beta_{3,2}$, $\beta_{4,1}$ and $\beta_{4,2}$, are significantly different from zero, unlike the average estimate of the inactive coefficients. A plot of the MCMC and variational posterior densities (the latter obtained in closed form), given in Figure \ref{FigPostDens} of Appendix \ref{AppSMCp}, shows that the posterior modes of the inactive coefficients are all zero. Moreover, in this case the variational distributions are usually solely made up of a clear spike at zero, whereas the MCMC histograms correspond roughly to a centered Gaussian distribution with average standard deviation $0.014$. Table $\ref{TabBetaOmega}$ also indicates a shrinkage effect for both variational and MCMC posterior means of the nonzero $\beta$ compared to the true values. This is a consequence of the spike-and-slab prior but does not seem to hamper the detection of the association signals, since the posterior probabilities of inclusion of the true nonzero associations are concentrated around $1$, while those corresponding to noise are usually much lower, whether obtained by MCMC, variational or simple Monte Carlo procedures; see Figure \ref{FigPPI} of Appendix \ref{AppSMCp}. 
Finally, the estimates of $\{\omega_s\}$ in Table $\ref{TabBetaOmega}$ provide a fair approximation to the actual proportion of responses associated with a given covariate. 

Two additional numerical experiments comparing variational and MCMC posterior quantities are provided in Appendix \ref{AppSMCp}. One compares the estimates of $\omega$ and $\tau$ with the true values when the data are generated from the model with $p=100$ covariates and $d=~10$ responses. It also provides receiver operating characteristic (ROC) curves assessing the pairwise variable selection performance for both inference types. The other simulation gathers the observed values, $y$, and the estimated posterior means of $X\beta$ obtained by variational and MCMC procedures and an oracle. Both experiments indicate equivalent performance for MCMC and variational inferences.

\section{Statistical performance}\label{SecNS}

\subsection{Predictor selection}\label{SecPred}

The problems considered in Section \ref{SecComp} were small enough to allow accurate and tractable MCMC inference. In this section, we assess the performance of our approach on larger problems by comparing it to popular variable selection methods; i.e., with joint modelling of outcomes and covariates (elastic net for multivariate Gaussian responses), with joint modelling of covariates only (Bayesian multiple regression based on MCMC inference, ``BAS'', or variational inference, ``varbvs''), or with fully marginal modelling (univariate ordinary least squares and ``lmBF'' Bayesian regressions). Complete descriptions and references are in Appendix \ref{AppMeths}. The methods are compared by measuring their ability to detect the active covariates, i.e., to determine which covariates are associated with at least one response. For our variational approach, this task is achieved by ranking the posterior means of the $\omega_s$, which control the proportion of responses associated with a given covariate. 

Our data-generation design is based on generally accepted principles of population genetics. We simulate SNPs under Hardy--Weinberg equilibrium from a binomial distribution with probabilities corresponding to minor allele frequencies of common variants, chosen in the interval $(0.05, 0.5)$ uniformly at random, and we generate outcomes from Gaussian distributions with specific error variances. The dependence structure of these variables is either enforced block-wise with preselected auto- or equicorrelation coefficients or chosen to be that of real data. The labels of the active SNPs and outcomes are picked randomly, and each active SNP is associated to one (randomly selected) active outcome and to each of the remaining active outcomes with a prescribed probability; some outcomes are therefore under pleiotropic control. The proportions of outcome variance explained per SNP are simulated for all associations from a positively skewed Beta distribution to favour the generation of smaller effects, and they are then rescaled to match a given average proportion. To mimic the result of natural selection, the effect sizes are inversely related to the SNP minor allele frequencies. For more details, see Appendix~\ref{AppGen}. 

 \begin{figure}
\centering
  \noindent\includegraphics[scale=0.72]{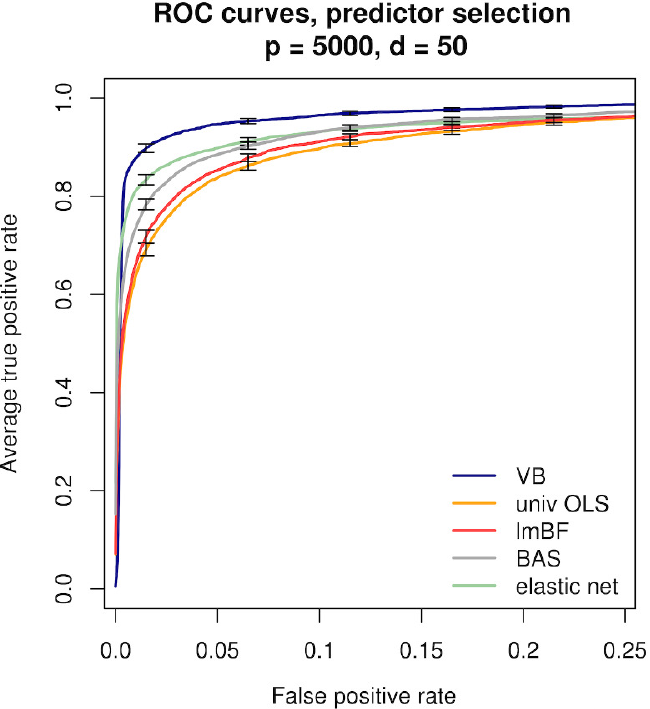}\quad
  \includegraphics[scale=0.72]{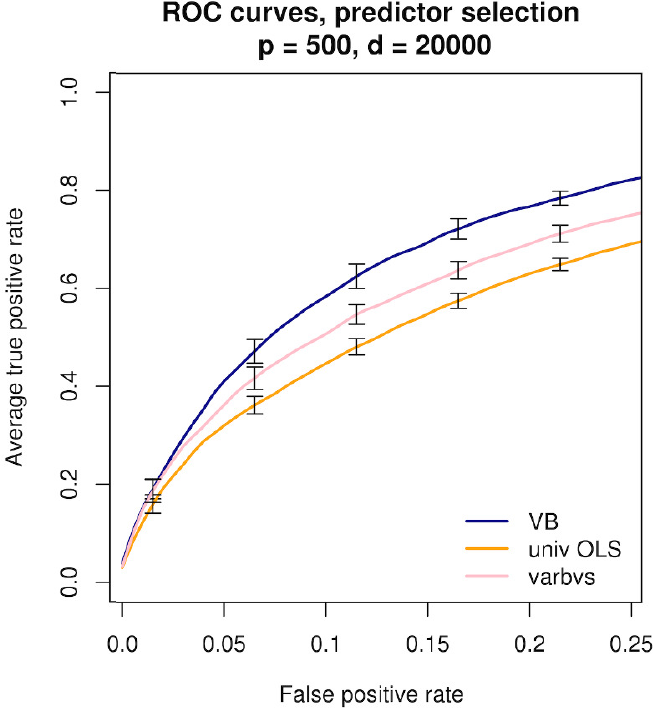}\quad
 \includegraphics[scale=0.72]{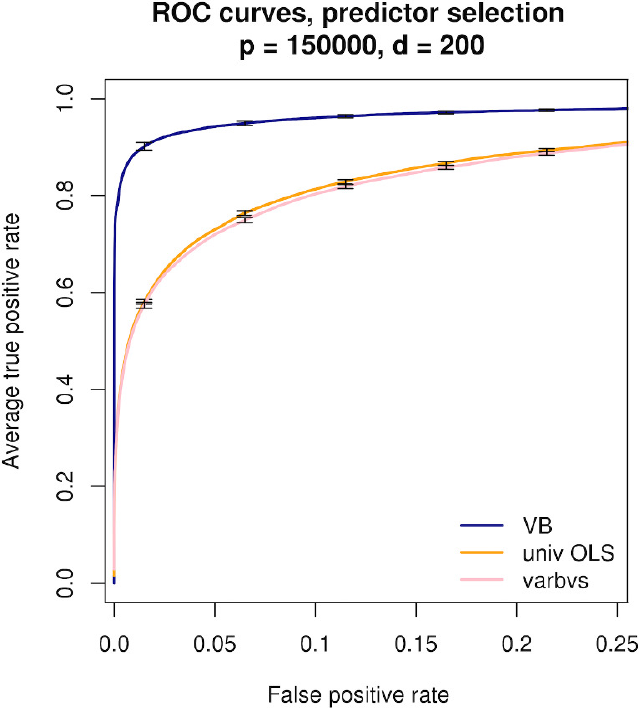}
    \caption{\small Truncated average receiver operating characteristic curves with $95\%$ confidence intervals for predictor selection obtained from $48$ replications. The competing methods are used in three studies of different sizes, based on their computational tractability. Left: $p=5,000$ covariates spatially autocorrelated with correlation coefficient $\rho_X = 0.75$, $d=50$ outcomes equicorrelated by blocks with four blocks of equal sizes and correlation coefficients $\rho_Y=0.8, 0.3, 0.2$ and $0.5$, $p_0=100$ active covariates, $d_0=40$ active outcomes, $n=250$ observations, probability of association with an additional outcome $p_{\text{add}} = 0.15$, average outcome variance percentage explained by the active covariates $p_{\text{ve}} = 30.0\%$. Middle: $p=500$ independent covariates, $d=20,000$ outcomes equicorrelated by blocks of size $10$ with $\rho_X \in \{0.5, \ldots, 0.8\}$, $p_0 = 300$, $d_0 = 12,500$, $n = 300$, $p_{\text{add}} = 0.01$, $p_{\text{ve}} = 55.8\%$. Right: $p=150,000$ covariates autocorrelated by blocks of size $100$ with $\rho_X \in \{0.5, \ldots, 0.9\}$, $d = 200$ outcomes with same correlation structure than real protein expression levels \citep[Diogenes study,][see Appendix \ref{AppPre}]{2010_Diogenes_Obesity_rev}, $p_0 = 500$, $d_0=150$, $n=200$, $p_{\text{add}} = 0.05$, $p_{\text{ve}} = 62.6\%$. The univariate ordinary least squares and varbvs curves overlap.
}\label{FigROCAll}
\end{figure} 

We perform $48$ replications for each of three simulation configurations. The first configuration has moderate numbers of covariates ($p=5,000$) and outcomes ($d=50$), and allows time-consuming methods to run within hours. The second has many outcomes ($d=20,000$) and the third has many covariates ($p=150,000$); these numbers approach those encountered in molecular QTL studies. The remaining settings (numbers of active outcomes and covariates, of observations, effect sizes, etc) are detailed in the caption to Figure \ref{FigROCAll}.

The ROC curves in Figure \ref{FigROCAll} indicate that our approach outperforms the other methods. It is appreciably more powerful for low false positive rates, which are of particular interest for the highly sparse scenarios typically expected for genome-wide association studies. Despite the correlation among the covariates and the outcomes, our method does not seem to suffer from the independence assumptions implied by the mean-field approximation, as suggested by the results of Section \ref{SecLB}. The marginal ordinary least squares and marginal lmBF regressions appear to miss many associations because of their univariate modelling of covariates, but jointly accounting for the covariates may not suffice, as suggested by the rather poor performances of the Bayesian multiple regression approaches, BAS and varbvs, which apply separate multiple linear regressions for each outcome. It appears that the ability of our approach to exploit the similarity across outcomes yields more power to detect their shared associations. Finally, even though the multivariate elastic net models jointly the covariate and outcome variables, its inference suffers from the assumption that to each covariate corresponds a single regression coefficient, shared for all responses. As a consequence, regression estimates of covariates with weak or few associations with the responses may be shrunk to zero.

\subsection{Combined selection of predictors and outcomes}

\begin{figure}
\centering
  \noindent \includegraphics[scale=0.97]{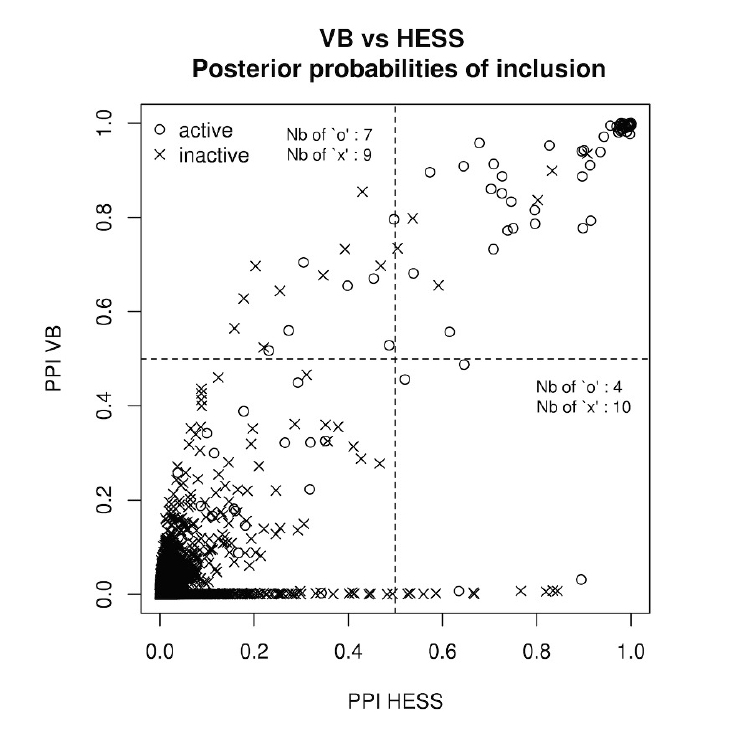} 
      \includegraphics[scale=0.97]{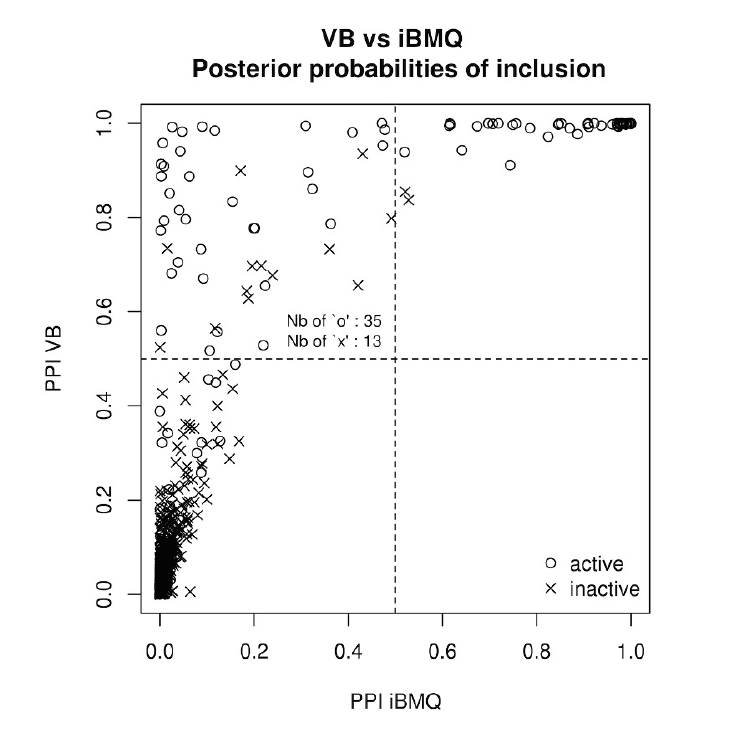} 
          \caption{
  \small Marginal posterior probabilities of inclusion (PPI) obtained by our approach, and those of HESS (left) and those of iBMQ (right), for a problem with $p=250$ covariates of which $p_0=50$ are active, with $d=100$ outcomes, of which $d_0= 50$ are active, and $n=250$ samples.}\label{FigVBHESSiBMQ}
\end{figure}

Unlike the classical variable selection methods used as comparators in Section \ref{SecPred}, our approach and those of \citet{richardson2010bayesian}, HESS, and \citet{scott2012integrated}, iBMQ, are tailored to molecular QTL problems: they quantify the associations between each covariate-response pair in a single model, and thus provide flexible and unified frameworks for detecting pairs of associated SNP-molecules, as well as pleiotropic SNPs associated with many molecular outcomes. 
In this section, we compare the three approaches in terms of the posterior quantities used to perform such selection. 
As both HESS and iBMQ rely on MCMC sampling, we consider smaller problems than in Section \ref{SecPred} in order to ensure convergence within a reasonable time. The simulated datasets have $p=250$ covariates, of which $p_0=50$ are active, and $d=100$ outcomes, of which $d_0=50$ are active, the probability of association being $0.05$, for $n=250$ samples. On average, the active covariates account for $22\%$ of the variance of an outcome with which they are associated. HESS was run with three MCMC chains, the number selected by the authors for their simulations but with $50,000$ iterations of which $25,000$ were discarded as burn-in. For iBMQ, $50,000$ iterations were saved after removal of $50,000$ burn-in samples, as suggested in the package documentation for a problem of comparable dimensions. Inference for one replication took on average $10$ seconds with our method, around $21$ minutes with iBMQ and $4$ hours with HESS (GPU computation option disabled, since no GPU was available to us) on an Intel Xeon CPU at 2.60 GHz with 64 GB RAM. 

\begin{table}
\begin{center}
\captionof{table}{\small Mean true positive rate (TPR) and true negative rate (TNR) for our approach, HESS and iBMQ based on median probability models. Settings: $p=250$, $p_0=50$, $d=100$, $d_0=50$, $n=250$, $48$ replicates. Standard errors are in parentheses.\label{TabTPR}}
\footnotesize
  \begin{tabular}{rcc}
  \hline
 $100 \times$ & TPR  & TNR \\ 
  \hline
  VB  & $58.9$ ($5.0$) & $99.9$ ($0.0$) \\ 
  HESS & $57.9$ ($5.3$) & $99.9$ ($0.0$)   \\ 
  iBMQ & $0.1$ ($0.2$) & $99.8$ ($0.0$)  \\ 
   \hline
\end{tabular}
\end{center}
\end{table}

Figure \ref{FigVBHESSiBMQ} compares the marginal posterior probabilities of inclusion obtained by our method with those of HESS and iBMQ. We observe a strong correlation between our approach and HESS, with a quite good ability to discriminate between active and inactive covariate-response pairs. There is a discrepancy at the zero ordinate, where HESS signals a series of false positives and few true positives. The comparison with iBMQ is more contrasted, as the values of its posterior probabilities of inclusion for many true associations are below $0.1$ and indistinguishable from noise. The same conclusions are reached when running the three methods on $47$ additional datasets, as suggested by Table \ref{TabTPR}, which gathers sensitivity and specificity measures based on median probability models \citep{barbieri2004optimal} (consisting of those covariate-response pairs whose posterior inclusion probability is higher than $0.5$).  As discussed in Section~\ref{SecMod}, control of signal sparsity can be induced through the prior for $\omega_s$, still, rather than median probability models, one may prefer to use a data-driven false discovery threshold in order to prescribe a desired level of false discoveries. 

\begin{figure}
\centering
    \noindent\includegraphics[scale=0.96]{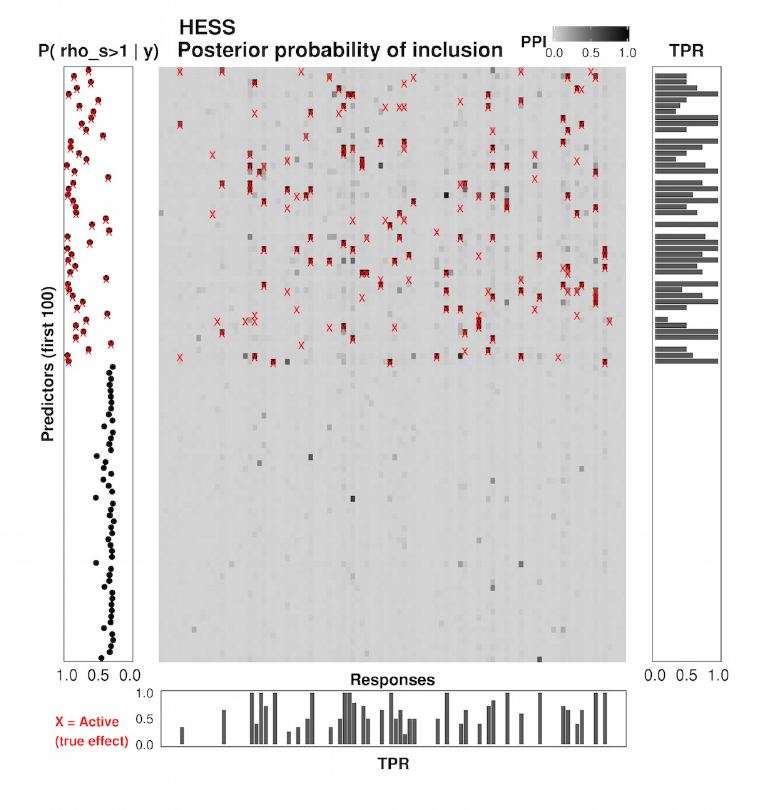} 
      \includegraphics[scale=0.96]{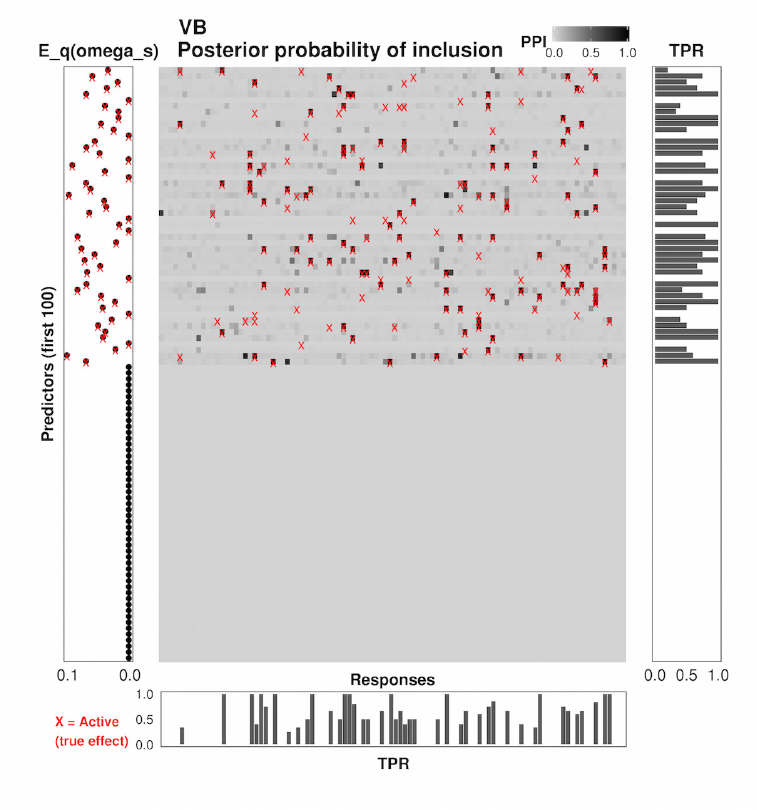} 
    \caption{\small Posterior quantities for detection of associations with HESS (left) and with our approach (right), for a simulated dataset with $p=250$ independent covariates (here only the first $100$ are displayed), of which $p_0=50$ are active and placed first, with $d=100$ responses ($50$ active) and $n=250$ individuals. Marginal posterior probabilities of inclusion (central panel), true positive rates for predictor and response selection based on posterior probability of inclusion being $> 0.5$ (bottom and right panels), posterior probability $\text{pr}(\rho_s > 1 \mid y)$ for HESS and posterior mean $\text{E}_q(\omega_s\mid y)$ for our approach (left panel). The simulated associations are shown by red crosses. \label{FigVBHESS}}
\end{figure}

Figure \ref{FigVBHESS} compares the patterns recovered by HESS and by our method, again based on the marginal posterior probabilities of inclusion from the first replicate. Visual comparison of the true positive rates suggests that the abilities of the two approaches to detect the true associations are very similar. Our approach indicates the presence of associations in the region of active covariates only, whereas the HESS pattern is blurrier in regions of inactive covariates. The posterior means of $\{\omega_s\}$ from our approach discriminate quite well between active and inactive covariates, and so do, for HESS, the posterior probabilities $\text{pr}(\rho_s >1 \mid y)$ ($s=1, \ldots, p$), described by \citet{richardson2010bayesian} as capturing the propensity for a given covariate to influence several responses simultaneously.

\subsection{Application to a real mQTL dataset}\label{SecReal}

We end these numerical experiments by illustrating our approach on data from a large multicenter dietary intervention study called Diogenes \citep{2010_Diogenes_Obesity_rev}. The study contains a series of genomic data types collected at different stages of a dietary treatment provided to the cohort. Its goal is to uncover molecular mechanisms underlying the metabolic status of overweight individuals and improve understanding of the factors predisposing weight regain after a diet. Here, we perform a metabolite quantitative trait locus (mQTL) analysis; in this context, the metabolites may be viewed as proxies for the clinical condition of interest, weight maintenance. We also use this illustration on real data to further highlight the benefits of modelling the outcomes jointly via an extensive permutation-based comparison with the single-response variational method varbvs \citep{carbonetto2012scalable}.

After quality control, the data consist of $p=215,907$ tag SNPs and $d=125$ metabolite expression levels, adjusted for age, center and gender, for $n=317$ individuals. The SNPs were collected on Illumina HumanCore arrays and the metabolites were quantified in plasma using liquid chromatography-mass spectrometry (LC-MS). They span cholesterol esters (CholE), phosphatidylcholines (PC), phosphatidylethanolamines (PE), sphingomyelins (SM), di- (DG) and triglycerides (TG). Appendix \ref{AppPre} provides more details.

\begin{table}
\begin{center}
\captionof{table}{\small Number of associations declared by our method and by varbvs, and number of signals in common at selected permutation-based false discovery rates. For each case, the number of associations also declared by univariate screening at Benjamini--Hochberg FDR of $25\%$ is in parentheses. \label{TabFDR}}
\footnotesize
\begin{tabular}{cccc}
  \hline 
  &\# declared:\\
Permutation-based FDR (\%) & VB & varbvs & VB $\cap$ varbvs \\ 
  \hline
  5 & 21 & 19 & 8 \\ 
  10 & 26 & 19 & 8 \\ 
  15 & 47 & 21 & 10 \\ 
  20 & 76 & 31 & 12 \\ 
  25 & 89 (48 univ.) & 47 (19 univ.) & 14 (13 univ.) \\ 
   \hline
\end{tabular}
\end{center}
\end{table}

\begin{figure}
\centering
  \noindent\includegraphics[scale=0.61]{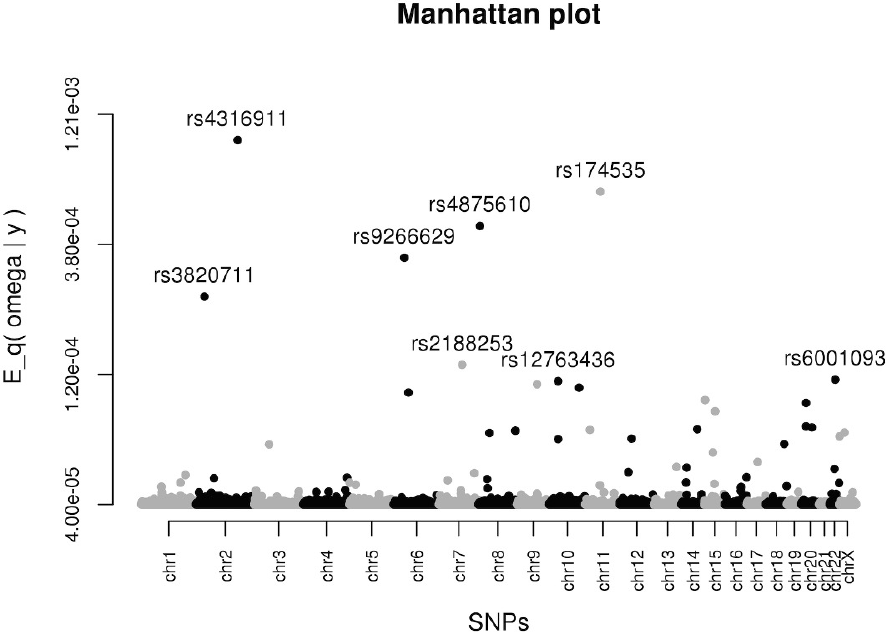}\quad 
    \includegraphics[scale=0.91]{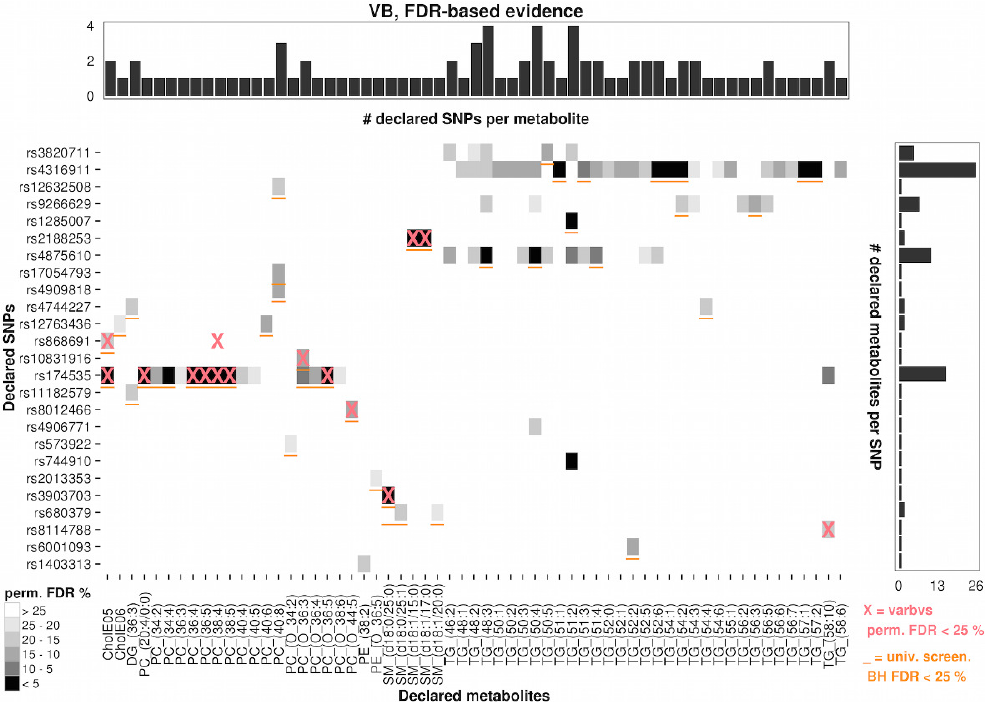} 
    \caption{\small SNPs and pairwise associations declared by our approach for the Diogenes study. Left: Manhattan plot for SNP association and evidence of pleiotropy. Right: posterior probabilities of inclusion for SNP-metabolite associations declared at estimated FDR of $25\%$ and overlap with the associations declared by the varbvs method at same FDR level (pink crosses) and declared by univariate screening at Benjamini--Hochberg FDR of $25 \%$ (orange underscores).}\label{FigReal}
\end{figure} 

In order to adjust for multiplicity, we specify the hyperparameters for $\omega$ according to the discussion of Section~\ref{SecMod} and choose the prior average number of active SNPs, $p^*$, by grid search within a $3$-fold cross-validation procedure that maximizes the variational lower bound. After hyperparameter selection, the algorithm converged in $83$ iterations, taking about $10$ hours on an Intel Xeon CPU at 2.60 GHz with 512 GB RAM. The posterior means $\text{E}_q(\omega_s \mid y)$ suggest the presence of several active SNPs, spread across the chromosomes (Figure \ref{FigReal}), but we use the marginal posterior inclusion probabilities, $\text{E}_q(\gamma_{st} \mid y)$, to declare pairwise associations and active SNPs. 

We compare varbvs and our method on real data based on the number of associations declared by each method at specific false discovery rates estimated by permutations. We apply Efron's Bayesian interpretation of the false discovery rate \citep{efron2008microarrays} to posterior probabilities of inclusion, and use an empirical null distribution based on $B=400$ permutations to compute the estimate 
\begin{equation}\label{EqFDR}\widehat{\textup{FDR}}(\tau) = \frac{\textup{median}_{b = 1, \ldots, B} \#\{\textup{PPI}_{st}^{(b)} > \tau\}}{\#\{\textup{PPI}_{st} > \tau\}}\,,\qquad 0 <\tau < 1\,,\end{equation}
for a grid of thresholds $\tau$; we then fit a cubic spline to the resulting false discovery rates to find thresholds for specific rates. The analysis suggests that our method is more powerful, with $89$ associations declared at an estimated FDR of $25 \%$, against $47$ for varbvs; the superiority of our method is further highlighted by Table $\ref{TabFDR}$. Figure~$\ref{FigReal}$ displays the associations declared by our method and the overlap with those declared by varbvs at estimated FDR of $25\%$; the associations detected also largely agree with those obtained with marginal screening at Benjamini--Hochberg FDR of $25\%$. 

Database searches on the functional relevance of the detected associations give hints of promising biological functions related to metabolic activities for $12$ of the $25$ SNPs declared as active by our procedure. For instance, the most outstanding SNP Figure \ref{FigReal}, $rs4316911$, shows many associations with triglyceride levels, and turns out to be located less than $150 kb$ from the protein coding gene ITGA6 known to be linked to diabetic kidney disease \citep{iyengar2015genome}. The second most prominent pleiotropic SNP, $rs174535$, is declared by our approach to be associated with phospholipids, more precisely with $14$ different phosphatidylcholine levels, of which four are ether-linked/plasmalogen (PC-O). Interestingly, this latter SNP has been recently reported to be related to metabolite levels; among others, it was found to be associated with trans fatty acid levels and plasma phospholipid levels \citep{mozaffarian2015genetic}, in line with our findings. Moreover, it was found to be an eQTL for the fatty acid desaturase genes FADS1 and FADS2. The SNP $rs3903703$ too has been identified as associated with very long-chain fatty acid levels \citep{lemaitre2015genetic}. This seems to agree with our findings, in which $rs3903703$ exhibits associations with sphingomyelin, a type of lipid containing fatty acids of different chain lengths. The complete subset of SNPs with metabolism-related links found by our procedure is given in Appendix \ref{AppE}.  Additional details on this real data study, as well as on its replication using simulated data, are also provided there. 

\section{Conclusions}\label{SecConcl}

We have described a scalable and efficient approach to joint variable selection from large numbers of candidate predictor and outcome variables. As it exploits the similarity across outcomes through a flexible hierarchical structure, our procedure outperforms the most popular predictor selection approaches in high-dimensional set-ups. The variational approximation on which our approach relies provides accurate posterior quantities, with reduced computational effort relative to MCMC procedures; in particular, it yields inferences comparable to those of the MCMC procedure HESS \citep{richardson2010bayesian}. Convergence control is automatic, whereas convergence assessment for MCMC algorithms can be difficult, especially in high dimensions. Our simulations also show that variable selection remains powerful when the predictors and outcomes are correlated, notwithstanding the independence assumptions underlying the mean-field factorization. 

The key added-value of our approach is its applicability to molecular QTL datasets without the need for prior dimension reduction. In an application, our approach recovered several previously reported SNP-metabolite associations, and declared more associations than the single-outcome method ``varbvs'' \citep{carbonetto2012scalable} at prescribed false discovery rates, thus highlighting the benefits of jointly modelling the outcomes. To the best of our knowledge, no competing Bayesian approach for joint inference on two high-dimensional sets of variables can deal with the problem sizes typically encountered in molecular QTL analyses. 

Bioinformatics is moving towards whole-genome analyses, for which several million genetic variants need to be considered, so it seems worthwhile to consider further speed-up strategies for our approach. One option is to use new optimization procedures. So-called natural gradient methods, which rely on the Riemannian structure of variational approximate distributions, seem particularly attractive, as they can be orders of magnitude faster than conventional gradient algorithms \citep{honkela2007natural}. Another possibility is a ``split-and-merge'' strategy, i.e., first partitioning the variable space and then inferring a global variational distribution on the aggregated dataset. \citet{tran2016parallel} designed a variant of this approach for sample space partitioning. At the recombination step, they proposed to ``merge'' the variational distributions by exploiting the independence assumptions of the mean-field formulation. Both strategies could lead to significant computational gains.

\section{Software}

The algorithm and the data-generation functions used in this paper are implemented in the publicly available R package \texttt{locus}.

\section*{Acknowledgements}
The authors are grateful to the Associate Editor and the two referees for their very helpful comments that improved the paper. We thank Jérôme Carayol, Loris Michel and Armand Valsesia for their valuable comments. We also thank James Holzwarth, Bruce O'Neel and Jaroslaw Szymczak for giving us access to computing resources. 

\section*{Funding}

The Diogenes trial was supported by the European Commission (FP6-2005-513946) and Nestlé Institute of Health Sciences.

\bibliographystyle{plainnat}
\bibliography{./refs}  

\appendix
\section{Multiplicity control at covariate level}\label{AppA}

Sparsity control at covariate level can be induced through the prior distribution of $\omega$, by carefully selecting its hyperparameters. The prior probability that $X_s$ is ``active'' (i.e., associated with at least one response) is
$$
p\left( \cup_{t=1}^d\{\gamma_{st}=1\} \right) = \int \left\{ 1- \prod_{t=1}^d p\left( \gamma_{st}=0 \mid \omega_s\right) \right\} p(\omega_s) \textup{d}\omega_s 
= 1 - \frac{\text{Beta}(a_s, b_s+d)}{\text{Beta}(a_s, b_s)}\,,
$$
and this after a little algebra equals
$$
1 - \frac{\prod_{j=1}^d (b_s + d -j )}{\prod_{j=1}^d ( a_s + b_s + d -j )}\,,
$$
so assuming exchangeability and setting 
\begin{equation}\label{SMEqHyper} 
a_s \equiv 1, \qquad \qquad b_s \equiv d (p-p^*)/p^*, \qquad\qquad 0 < p^* < p,
\end{equation} 
implies that
$$
p\left( \cup_{t=1}^d\{\gamma_{st}=1\} \right) = \frac{d}{b_s+d} = \frac{p^*}{p}\,,
$$
where $p^*$ is interpreted as a prior average number of active covariates. The choice (\ref{SMEqHyper}) yields a multiplicity adjustment as suggested by a plot (Figure \ref{FigPOR}) of the prior odds ratios, indicating the penalty induced by the prior when moving from $q_s-1$ to $q_s$ responses associated with $X_s$, 
\begin{eqnarray*}
\text{POR}(q_s-1:q_s)&=&\frac{p\left( \sum_{t=1}^d\gamma_{st}=q_s-1\right)}{p\left( \sum_{t=1}^d\gamma_{st}=q_s\right)} 
= \frac{\text{Beta}(a_s+q_s-1, b_s+d-q_s+1)}{\text{Beta}(a_s+q_s, b_s+d-q_s)}\\
 &=&\frac{b_s + d - q_s }{ a_s + q_s -1 }\,, \hspace{7cm}  q_s=1, \ldots, d\,.
\end{eqnarray*}

\section{Derivation of the variational algorithm}\label{AppB}

\subsection{Variational distributions}
We provide the detailed derivation of our variational algorithm, which is given in Appendix \ref{AppAlg}. We have
\begin{eqnarray}\label{SMEqTrue}
p\left(y, \beta, \gamma, \tau, \sigma^{-2}, \omega\right) &=& 
p\left(y \mid \beta, \tau \right) \,p\left(\beta \mid \gamma, \tau, \sigma^{-2} \right) \,p\left(\gamma \mid\omega)\,  p( \omega)\,p(\tau) \,p(\sigma^{-2}\right) \nonumber \\
&=& \left\{\prod_{t=1}^dp\left(y_t \mid \beta_t, \tau_t \right)\right\} \left\{\prod_{s=1}^p\prod_{t=1}^d p\left(\beta_{st} \mid \gamma_{st}, \tau_t, \sigma^{-2} \right) \right\} \left\{\prod_{s=1}^p\prod_{t=1}^d p\left(\gamma_{st}\mid \omega_s\right) \right\}\nonumber\\ 
&&\times\left\{\prod_{s=1}^pp(\omega_s)\right\}\left\{\prod_{t=1}^dp(\tau_t)\right\}p\left(\sigma^{-2}\right)\,,
\end{eqnarray}
where 
\begin{eqnarray*}
p(y_t \mid \beta_t, \tau_t) 
&=& (2\pi)^{-n/2}(\tau_t)^{n/2} \exp\left(-\frac{\tau_t}{2} \Vert y_t - X \beta_t\Vert^2\right)\,, \qquad y_t \in \mathbb{R}\,,\\
p(\beta_{st} \mid \gamma_{st}, \tau_t, \sigma^{-2}) &=& \left\{(2\pi)^{-1/2}(\sigma^{-2}\tau_t)^{1/2} \exp\left(-\frac{\sigma^{-2}\tau_t}{2} \beta_{st}^2\right) \right\}^{\gamma_{st}} \delta_0(\beta_{st})^{1-\gamma_{st}}\,, \qquad \beta_{st} \in \mathbb{R}\,,\\
p(\gamma_{st}\mid \omega_s) &=& \omega_s^{\gamma_{st}} \left(1-\omega_s\right)^{1-\gamma_{st}}\,,\qquad \gamma_{st} = 0, 1,\\
p(\omega_s) &=& \omega_s^{a_s-1} (1-\omega_s)^{b_s-1}/ \text{B}(a_s, b_s)\,, \qquad 0 < \omega_s < 1\,,\\
p(\tau_t) &=&\tau_t^{\eta_t-1}\exp\left( -\kappa_t\tau_t\right) \kappa_t^{\eta_t}/\Gamma(\eta_t)\,,\qquad \tau_t > 0\,,\\
p(\sigma^{-2}) &=&\left(\sigma^{-2}\right)^{\lambda-1}\exp\left( -\nu\sigma^{-2}\right) \nu^{\lambda}/\Gamma(\lambda)\,, \qquad \sigma^2 > 0\,.
\end{eqnarray*}
Let $\theta=\left(\beta, \gamma, \tau, \sigma^{-2}, \omega\right)$ and consider the following \emph{mean-field} form for the variational approximation,
$$
q\left(\theta\right) =  \left\{  \prod_{s=1}^p\prod_{t=1}^d q(\beta_{st}, \gamma_{st}) \right\}\left\{\prod_{s=1}^p q(\omega_s)\right\}\left\{\prod_{t=1}^{d} q(\tau_t) \right\} q\left(\sigma^{-2}\right)\,.
$$
We obtain each component of this factorization using the formula 
$$
\log q_j(\theta_j) = \text{E}_{-j} \{\log p(y, \theta)\} + \text{cst},\qquad\qquad j=1, \ldots, J,
$$
with $p(y, \theta)$ given in $(\ref{SMEqTrue})$, where $\text{E}_{-j}$ is the expectation with respect to the distribution $q_k$ over all variables $\theta_k$ ($k\neq j$) and where $\text{cst}$ is a constant with respect to $\theta_j$.  
Writing $\theta^{(r)}_j$ for the $r^{th}$ moment with respect to the approximate posterior distribution $q_j$ of $\theta_j$, we have
\begin{eqnarray*}
\log q(\beta_{st}, \gamma_{st}) &=& \sum_{k=1}^d\text{E}_{-(\beta_{st}, \gamma_{st})}\left\{\log p(y_k \mid \beta_k, \tau_k)\right\} + \sum_{j=1}^p \sum_{k=1}^d \text{E}_{-(\beta_{st}, \gamma_{st})}\left\{\log p(\beta_{jk} \mid \gamma_{jk}, \tau_k, \sigma^{-2})\right\}\\
&&+ \sum_{j=1}^p \sum_{k=1}^d \text{E}_{-(\beta_{st}, \gamma_{st})}\left\{\log p(\gamma_{jk}\mid \omega_j)\right\} + \text{cst}\\
&=& -\frac{1}{2}\tau_t^{(1)}\text{E}_{-\beta_{st}}\Vert y_t - X \beta_t \Vert^2 + \gamma_{st}\frac{1}{2}\left\{ \text{E}\left( \log \sigma^{-2} \right)+ \text{E} \left(\log \tau_t \right)- \log(2\pi)\right\}\\&&- \gamma_{st}\frac{\left(\sigma^{-2}\right)^{(1)} \tau_t^{(1)}}{2}\beta_{st}^2+(1-\gamma_{st})\delta_0(\beta_{st}) +\gamma_{st}\text{E}\left(\log \omega_s\right) + (1-\gamma_{st})\text{E}\left\{\log(1-\omega_s)\right\}+ \text{cst} \\
&=& -\gamma_{st}\frac{\tau_t^{(1)}}{2}\left[\beta_{st}^2\left\{\Vert X_{s}\Vert^2 +\left(\sigma^{-2}\right)^{(1)}\right\}-2\beta_{st} \left\{y_{t}^TX_{s}-X_{s}^T\sum_{j=1, j\neq s}^p\beta_{jt}^{(1)}X_{j}\right\} \right]\\ &&+\gamma_{st}\frac{1}{2}\left\{ \text{E}\left(\log \sigma^{-2}  \right)+ \text{E} \left(\log \tau_t \right)- \log(2\pi)\right\}+(1-\gamma_{st}) \delta_0(\beta_{st})+\gamma_{st}\text{E}\left(\log \omega_s \right)\\ &&+ (1-\gamma_{st})\text{E}\left\{\log(1-\omega_s)\right\}+ \text{cst}\,,
\end{eqnarray*}
where $\text{cst}$ is a constant with respect to $\beta_{st}$ and $\gamma_{st}$. Completing the square yields
\begin{eqnarray*}
q(\beta_{st}, \gamma_{st}) &=& \text{cst} \left[\left(2\pi \sigma_{\beta, st}^2\right)^{-1/2}\exp\left\{-\frac{1}{2\sigma_{\beta, st}^2}\left( \beta_{st}-\mu_{\beta, st}\right)^2 \right\}\right]^{\gamma_{st}} \\
&&\times\left[\left[\exp\left\{ \text{E}\left(\log \sigma^{-2}  \right)+ \text{E} \left(\log \tau_t \right)\right\}\sigma^2_{\beta, st}\right]^{1/2}\exp\left(\frac{1}{2}\mu^2_{\beta, st}\sigma^{-2}_{\beta, st}\right) \exp\left\{\text{E}\left(\log \omega_s \right)\right\} \right]^{\gamma_{st}}\\ 
&&\times\left\{\delta_0(\beta_{st})\right\}^{1-\gamma_{st}} \exp\left[ \text{E}\left\{\log\left(1-\omega_s\right)\right\}\right]^{1-\gamma_{st}}\,,
\end{eqnarray*}
with 
$$
\mu_{\beta, st} = \sigma_{\beta, st}^2\tau_t^{(1)}\left\{y_{t}^TX_{s}-X_{s}^T\sum_{j=1, j\neq s}^p\mu_{\beta, jt}\gamma_{jt}^{(1)} X_{j}\right\}\,,\qquad\quad
\sigma^2_{\beta, st} = 
 \frac{1}{\tau_t^{(1)}\left\{\Vert X_{s}\Vert^2 + \left(\sigma^{-2}\right)^{(1)}\right\}}\,.
$$
We therefore observe that 
$$
q(\beta_{st}, \gamma_{st}) = q(\beta_{st} \mid \gamma_{st}) q(\gamma_{st})\,,
$$ 
with 
$$
\beta_{st} \mid \gamma_{st}=1, y\; \sim \mathcal{N}\left(\mu_{\beta, st}, \sigma^2_{\beta, st}\right)\,,\quad
\beta_{st} \mid \gamma_{st}=0, y\; \sim \delta_0\,,\quad
\gamma_{st} \mid y \sim \text{Bernoulli}\left(\gamma^{(1)}_{st}\right)\,,
$$
and with
$$
\frac{\gamma^{(1)}_{st}}{1-\gamma^{(1)}_{st}} 
= \sigma_{\beta, st}\exp\left[ \text{E}\left(\log \omega_s \right) - \text{E}\left\{\log\left(1-\omega_s\right)\right\} + \frac{1}{2} \left\{\text{E}\left(\log \tau_t \right)+\text{E}\left(\log \sigma^{-2} \right)\right\} + \frac{1}{2}\mu^2_{\beta, st}\sigma^{-2}_{\beta, st} \right]\,,
$$
i.e.,
 \begin{eqnarray}\label{SMEqEGam}
\gamma^{(1)}_{st} = \left[1 + \sigma_{\beta, st}^{-1} \exp\left\{ \text{E}\left\{\log\left(1-\omega_s\right)\right\} - \text{E}\left(\log \omega_s \right)  -\frac{1}{2} \text{E}\left(\log \tau_t \right) -\frac{1}{2} \text{E}\left(\log \sigma^{-2} \right) - \frac{1}{2}\mu^2_{\beta, st}\sigma^{-2}_{\beta, st} \right\}\right]^{-1}. \qquad
\end{eqnarray}
We now compute the variational approximate distribution for the error variance of each $y_t$:
\begin{eqnarray*}
\log q(\tau_t) &=& \text{E}_{-\tau_t}\left\{ \log p\left(y_t\mid \beta_t, \tau_t \right)\right\} +\sum_{s=1}^p  \text{E}_{-\tau_t}\left\{ \log p\left( \beta_{st} \mid \gamma_{st}, \tau_t, \sigma^{-2} \right) \right\} + \log p(\tau_t)+ \text{cst}\\
&=& \frac{n}{2}\log \tau_t  - \frac{\tau_t}{2}\,\text{E}\left(\Vert y_t - X\beta_t \Vert^2\right)+\frac{1}{2}\log \tau_t \sum_{s=1}^p \gamma_{st}^{(1)}- \frac{\tau_t}{2}\left(\sigma^{-2}\right)^{(1)}\sum_{s = 1}^p\beta_{st}^{(2)}+ (\eta_t-1)\log \tau_t \\
&&-\kappa_t\tau_t + \text{cst}\\
&=&\log \tau_t  \left(  \eta_t  + \frac{n}{2} + \frac{1}{2}\sum_{s=1}^p\gamma_{st}^{(1)} -1 \right)-\tau_t\left[ \kappa_t + \frac{1}{2} \Vert y_{t}\Vert^2- y_{t}^T\sum_{s=1}^p \mu_{\beta, st} \gamma^{(1)}_{st} X_{s}\right.\\
&&\left. +\sum_{s=1}^{p-1}\mu_{\beta, st}\gamma^{(1)}_{st} X_{s}^T\sum_{j=s+1}^p\mu_{\beta, jt}\gamma^{(1)}_{jt}X_{j}+ \frac{1}{2}\sum_{s=1}^p  \gamma_{st}^{(1)}\left(\sigma^2_{\beta,st} + \mu^2_{\beta, st} \right) \left\{\Vert X_{s}\Vert^2+\left(\sigma^{-2}\right)^{(1)} \right\}\right]+ \text{cst}\,.
\end{eqnarray*}
Therefore we have
$$\tau_t \mid y \sim \text{Gamma}\left(\eta_t^*, \kappa_t^*\right)$$
and
$$
\tau_t^{(1)}= \eta_t^* /\kappa_t^*\,,
$$
where
\begin{eqnarray*}
\eta^*_t &=& \eta_t+\frac{n}{2}+\frac{1}{2}\sum_{s=1}^p\gamma_{st}^{(1)}\,,\\
\kappa^*_t &=& \kappa_t+\frac{1}{2} \Vert y_{t}\Vert ^2- y_{t}^T\sum_{s=1}^p \mu_{\beta, st} \gamma^{(1)}_{st} X_{s} + \sum_{s=1}^{p-1}\mu_{\beta, st}\gamma^{(1)}_{st} X_{s}^T\sum_{j=s+1}^p\mu_{\beta, jt}\gamma^{(1)}_{jt}X_{j}\\
&&+ \frac{1}{2}\sum_{s=1}^p  \gamma_{st}^{(1)}\left(\sigma^2_{\beta,st} + \mu^2_{\beta, st} \right) \left\{ \Vert X_{s}\Vert^2+\left(\sigma^{-2}\right)^{(1)} \right\}\,.
\end{eqnarray*}
Since $\tau_t$ has a Gamma distribution, the expectation $\text{E}\left(\log \tau_t \right)$ appearing in $(\ref{SMEqEGam})$ can be rewritten in terms of $\eta_t^*$ and $\kappa_t^*$ using the digamma function \citep{miller1968handbook},
$$\Psi(x) = \frac{\text{d}}{\text{d}x}\log \Gamma(x) = \frac{\Gamma'(x)}{\Gamma(x)}\,,$$
as
\begin{equation}\label{SMEqETau}
\text{E}_q\left(\log \tau_t \right) =\Psi(\eta^*_t)-\log(\kappa^*_t)\,.\end{equation}
We also find that
\begin{eqnarray*}
\log q\left(\sigma^{-2}\right) &=&\sum_{s=1}^p \sum_{t=1}^d  \text{E}_{-\sigma^{-2}}\left\{ \log p\left( \beta_{st} \mid \gamma_{st}, \tau_t, \sigma^{-2} \right) \right\} + \log p(\sigma^{-2}) + \text{cst}\\
&=&\left( \frac{1}{2}\sum_{s=1}^p\sum_{t=1}^d\gamma^{(1)}_{st}\right)\log \sigma^{-2} -\sigma^{-2}\sum_{t=1}^d\frac{\tau^{(1)}_t}{2}\sum_{s=1}^p \beta_{st}^{(2)}+\left(\lambda -1 \right) \log \sigma^{-2} -\nu\sigma^{-2}+\text{cst}\\
&=& \left(\lambda + \frac{1}{2}\sum_{s=1}^p\sum_{t=1}^d\gamma^{(1)}_{st} - 1\right) \log \sigma^{-2}-\sigma^{-2}\left\{\nu + \sum_{t=1}^d\frac{\tau^{(1)}_t}{2}\sum_{s=1}^p \left(\sigma^2_{\beta,st} + \mu^2_{\beta, st} \right) \gamma_{st}^{(1)}\right\}+\text{cst}\,.
\end{eqnarray*}
Thus
$$\sigma^{-2}\mid y\sim \text{Gamma}\left(\lambda^*, \nu^*\right)\,,$$
$$
\left(\sigma^{-2}\right)^{(1)}= \lambda^* / \nu^*\,,$$
where
$$
\lambda^*=\lambda+\frac{1}{2}\sum_{s=1}^p\sum_{t=1}^d\gamma^{(1)}_{st}\,,\qquad
\nu^* = \nu+\frac{1}{2}\sum_{t=1}^d\tau^{(1)}_t\sum_{s=1}^p \left(\sigma^2_{\beta,st} + \mu^2_{\beta, st} \right) \gamma_{st}^{(1)}\,,
$$
and, as before, we now have
\begin{equation}\label{digammasigma}\text{E}_q\left(\log \sigma^{-2} \right) = \Psi(\lambda^*)-\log \nu^* \,. \end{equation}
Finally, we have
  \begin{eqnarray*}
 \log q(\omega_s) &=&\sum_{t=1}^d\text{E}_{\gamma_{st}}\left\{\log p(\gamma_{st} \mid \omega_s) \right\} + \log p(\omega_s)+\text{cst} \\ 
 &=&\left(\sum_{t=1}^d\gamma_{st}^{(1)}\right) \log \omega_s  + \left\{\sum_{t=1}^d\left(1-\gamma^{(1)}_{st}\right)\right\}\log(1-\omega_s)+ (a_s-1)\log \omega_s+ (b_s-1)\log(1-\omega_s) +\text{cst}\\
 &=&  \left(a_s+ \sum_{t=1}^d\gamma_{st}^{(1)} -1 \right) \log \omega_s+  \left(b_s-\sum_{t=1}^d\gamma_{st}^{(1)}+d-1\right)\log\left(1-\omega_s\right)+\text{cst}\,,
 \end{eqnarray*}
 that is,
 $$\omega_s\mid y\sim \text{Beta}\left( a^*_s,\, b_s^*\right)\,,$$
 and 
$$
\omega_s^{(1)}= \frac{a_s^*}{a_s^*+b_s^*}\,, 
$$
where 
$$a_s^*=a_s + \sum_{t=1}^d\gamma_{st}^{(1)}\qquad \qquad\qquad b_s^*= b_s-\sum_{t=1}^d\gamma_{st}^{(1)}+d\,. $$

\noindent As $\omega_s$ has a Beta distribution, we also get \citep{miller1968handbook}
\begin{eqnarray}\label{expect}
\text{E}_q\left(\log \omega_s \right) &=& \Psi(a_s^*) - \Psi(a_s^*+b_s^*) = \Psi\left(a_s + \sum_{t=1}^d\gamma_{st}^{(1)}\right) - \Psi\left(a_s + b_s+d\right)  \,,\nonumber\\
 \text{E}_q\left\{\log(1-\omega_s)\right\} &=& \Psi(b_s^*) - \Psi(a_s^*+b_s^*)= \Psi\left(b_s-\sum_{t=1}^d\gamma_{st}^{(1)}+d\right) - \Psi\left(a_s + b_s+d\right)\,. \end{eqnarray}

\subsection{Lower bound of the marginal log-likelihood}\label{AppLB}

\noindent We provide the computational details for the lower bound, $\mathcal{L}(q)$, of the marginal log-likelihood, $\log p(y)$. It is evaluated at each iteration of our algorithm, in order to monitor its convergence:
\begin{eqnarray*}
\mathcal{L}(q)&=& \int q(z)\log\left\{\frac{p(y, z)}{q(z)}\right\}\text{d}z \\
&=&\sum_{t=1}^dA\left(y_t \mid \beta_t, \tau_t \right) + \sum_{s=1}^p\sum_{t=1}^d B\left( \beta_{st}, \gamma_{st}\mid \tau_t, \sigma^{-2}\right) + \sum_{t=1}^d C(\tau_t) + D\left(\sigma^{-2}\right) + \sum_{s=1}^pG(\omega_s)\,, 
\end{eqnarray*}
with
\begin{eqnarray*}
A\left(y_t \mid \beta_t, \tau_t\right) &=&\text{E}_q\left\{\log p(y_t \mid \beta_t, \tau_t)\right\}
=-\frac{n}{2} \log(2\pi)+\frac{n}{2}\text{E}\left(\log \tau_t \right)-\frac{1}{2}\tau_t^{(1)}\text{E}\left\{\Vert y_t - X \beta_t \Vert^2\right\} \\
&=&-\frac{n}{2} \log(2\pi)+\frac{n}{2}\text{E}\left(\log \tau_t \right)
-\frac{1}{2}\tau_t^{(1)}\left\{ \Vert y_{t}\Vert^2-2y_{t}^T \sum_{s=1}^p\mu_{\beta, st}\gamma_{st}^{(1)}X_{s}  \right.\\
&&\left.+ 2\sum_{s=1}^{p-1}\mu_{\beta, st}\gamma_{st}^{(1)} X_{s}^T \sum_{j=s+1}^p \mu_{\beta, jt}\gamma_{jt}^{(1)} X_{j}+\sum_{s=1}^p\Vert X_{s}\Vert^2 \left(\sigma^2_{\beta, st} + \mu_{\beta, st}^2\right)\gamma_{st}^{(1)}\right\} \\
&=&-\frac{n}{2} \log(2\pi)+\frac{n}{2}\text{E}\left(\log \tau_t \right)-\tau_t^{(1)}\left\{\kappa_t^*-\frac{1}{2}\sum_{s=1}^p\gamma_{st}^{(1)} \left(\sigma_{\beta, st}^2+\mu_{\beta, st}^2\right)\left(\sigma^{-2}\right)^{(1)}-\kappa_t\right\}\,,
\end{eqnarray*}
\begin{eqnarray*}
B\left( \beta_{st}, \gamma_{st}\mid \tau_t, \sigma^{-2}\right) &=& \text{E}_q\left\{\log p(\beta_{st} \mid \gamma_{st}, \tau_t, \sigma^{-2})\right\} +  \text{E}_q\left\{\log p( \gamma_{st}\mid \omega_s)\right\} -  \text{E}_q\left\{\log q(\beta_{st}, \gamma_{st})\right\}\\
&=& \frac{1}{2} \gamma_{st}^{(1)} \left\{ -\log(2\pi) + \text{E} \left(\log \sigma^{-2} \right)  + \text{E} \left(\log \tau_t \right) - \left(\sigma^{-2}\right)^{(1)} \tau_t^{(1)}\left(\sigma^2_{\beta, st} + \mu_{\beta, st}^2\right)\right\}\\
&&+ \text{E}_q\left\{\left(1-\gamma_{st}\right)\,\delta_0(\beta_{st})\right\} + \gamma_{st}^{(1)} \text{E}\left(\log \omega_s \right)+ \left(1-\gamma_{st}^{(1)}\right) \text{E}\left\{\log(1-\omega_s)\right\}\\
&&+\frac{1}{2}\gamma_{st}^{(1)} \left\{\log(2\pi) +\log \sigma^2_{\beta, st}\right\} +\frac{1}{2\sigma_{\beta,st}^2}\text{E}_q\left\{\gamma_{st}\left(\beta_{st}-\mu_{\beta, st}\right)^2\right\}-\gamma_{st}^{(1)}\log \gamma_{st}^{(1)} \\
&&-\text{E}_q\left\{\left(1-\gamma_{st}\right)\,\delta_0\left(\beta_{st}\right) \right\} -\left(1-\gamma_{st}^{(1)}\right)\log\left(1-\gamma_{st}^{(1)}\right)\\
&=& \frac{1}{2} \gamma_{st}^{(1)} \left\{  \text{E} \left(\log \sigma^{-2} \right)  + \text{E} \left(\log \tau_t \right) \right\}-\frac{1}{2}\left(\sigma^{-2}\right)^{(1)} \tau_t^{(1)} \gamma_{st}^{(1)} \left(\sigma^2_{\beta, st} + \mu_{\beta, st}^2\right)\\
&&+ \gamma_{st}^{(1)}\text{E}\left(\log \omega_s \right)+ \left(1-\gamma_{st}^{(1)}\right)\text{E}\left\{\log(1-\omega_s)\right\}+\frac{1}{2}\gamma_{st}^{(1)} \left(\log \sigma^2_{\beta, st} +1\right)\\
&&
-\gamma_{st}^{(1)}\log \gamma_{st}^{(1)} -\left(1-\gamma_{st}^{(1)}\right)\log\left(1-\gamma_{st}^{(1)}\right)\,,
\end{eqnarray*}
\begin{eqnarray*}
C(\tau_t) &=& \text{E}_q\left\{\log p(\tau_{t})\right\} - \text{E}_q\left\{\log q(\tau_{t})\right\} \\
 &=&\left(\eta_t-\eta_t^*\right)\text{E}\left(\log \tau_t \right)-\left(\kappa_t-\kappa_t^*\right)\tau_t^{(1)}+\eta_t\log \kappa_t - \eta^*_t\log \kappa_t^* -\log \Gamma(\eta_t) +\log \Gamma(\eta_t^*) \,,
 \end{eqnarray*}
\begin{eqnarray*}
D\left(\sigma^{-2}\right)&=& \text{E}_q\left\{\log p\left(\sigma^{-2}\right)\right\}- \text{E}_q\left\{\log q\left(\sigma^{-2}\right)\right\} \\
&=&\left(\lambda-\lambda^*\right)\text{E} \left(\log \sigma^{-2} \right) -\left( \nu- \nu^*\right) \left(\sigma^{-2}\right)^{(1)} + \lambda \log \nu  -\lambda^*\log \nu^* - \log \Gamma\left(\lambda\right) +\log \Gamma\left(\lambda^*\right) \,,
\end{eqnarray*}
\begin{eqnarray*}
 G(\omega_s) &=&\text{E}_q\left\{\log p(\omega_s)\right\}-\text{E}_q\left\{\log q(\omega_s)\right\} \\
  &=& \left(a_s-a_s^*\right)\text{E}\left(\log \omega_s \right)+\left(b_s-b_s^*\right)\text{E}\left\{\log(1-\omega_s)\right\}
-\log B(a_s, b_s) +\log B\left(a_s^*, b_s^*\right) \,,
\end{eqnarray*}
where $\text{E}_q\left(\log \tau_t \right)$, $\text{E}_q\left(\log \sigma^{-2} \right)$, $\text{E}_q\left(\log \omega_s \right)$ and $\text{E}_q\left\{\log(1-\omega_s)\right\}$ are given by $(\ref{SMEqETau})$, $(\ref{digammasigma})$ and~$(\ref{expect})$.
\newpage
\subsection{Variational algorithm}\label{AppAlg}
$\text{}$\vspace{-0.1cm}\\
\begin{algorithm}[!ht]
{\small
  \caption{}
  \begin{algorithmic}
\Inputs{$y$ (centered), $X$ (standardized using the usual unbiased estimator of the variance), $a$, $b$, $\eta$, $\kappa$, $\lambda$, $\nu$, \text{tol}, \text{maxit}}\\
    \Initialize{$M=\{\mu_{\beta, st}\}$, $\Sigma=\left\{\sigma^2_{\beta, st}\right\}$, $\Gamma^{(1)}=\left\{\gamma^{(1)}_{st}\right\}$, $\tau^{(1)}=\left\{\tau_t^{(1)}\right\}$\\ $\mathcal{L}(q) \gets - \infty$, $\text{it} \gets 0$}\\
       \Repeat{\bf :}
      \State $\left(\sigma^{-2}\right)^{(1)} \gets \lambda^*/ \nu^*,$ \quad where\; $\lambda^*=\lambda + \frac{1}{2}\mathds{1}_p^T\,\Gamma^{(1)}\,\mathds{1}_d$, 
       \State\hspace{4.cm} $\nu^* = \nu + \frac{1}{2} \mathds{1}_p^T \left\{ (\Sigma + M \odot M ) \odot \Gamma^{(1)}\right\} \tau^{(1)}$
      \Comment{$\boldsymbol{\text{\bf E}_q\left( \sigma^{-2}\mid y\right)}$}
      \State $\tau^{(1)} \gets \eta^* \oslash \kappa^*,$  \quad where\; $\eta^* = \eta + \frac{n}{2}\mathds{1}_d + \frac{1}{2} \left(\Gamma^{(1)}\right)^{T} \mathds{1}_p,$\; 
      \State \hspace{3.65cm}$\kappa^*= \kappa + \frac{1}{2} \left(y \odot y\right)^T \mathds{1}_n - \left\{ X \left(M \odot \Gamma^{(1)}\right) \odot y\right\}^T \mathds{1}_n$ 
      \State \hspace{4.3cm} $+\left(\sum_{s=1}^{p-1} A_s \odot \sum_{j=s+1}^p A_{j}\right)^T\mathds{1}_n$
     \State \hspace{4.3cm} $+ \frac{1}{2} \left\{ n-1  
     + \left(\sigma^{-2}\right)^{(1)}\right\} \left\{ \Gamma^{(1)} \odot \left( \Sigma + M \odot M\right)\right\}^T \mathds{1}_p,$ 
     \State \hspace{3.5cm} $A_s= X_s\left(M\odot \Gamma^{(1)}\right)_{s\cdot}$
      \Comment{$\boldsymbol{\text{\bf E}_q\left( \tau_t \mid y\right)}$}
      \State $\Sigma \gets \mathds{1}_p \mathds{1}_d^T \oslash B,$\quad where $B= \left\{ n-1  
      + \left(\sigma^{-2}\right)^{(1)}\right\}\mathds{1}_p\left(\tau^{(1)}\right)^T$\;  
      \Comment{$\boldsymbol{\text{\bf Var}_q\left( \beta_{st} \mid \gamma_{st}=1, y\right)}$}
       \State $\log(\tau)^{(1)} \gets \Psi(\eta^*) - \log(\kappa^*)$
       \State $\log\left(\sigma^{-2}\right)^{(1)} \gets \Psi(\lambda^*) - \log(\nu^*)$
      \State $\log(\omega)^{(1)} \gets \Psi\left(a + \Gamma^{(1)}\mathds{1}_d\right) - \Psi\left(a+b+d \mathds{1}_p\right)$
      \State $\log(1-\omega)^{(1)} \gets \Psi\left(b - \Gamma^{(1)}\mathds{1}_d + d \mathds{1}_p\right) - \Psi\left(a+b+d \mathds{1}_p\right)$ 
      \For{$s = 1, \ldots, p$}
       \For{$t = 1, \ldots, d$}
      \State $M_{st} \gets \Sigma_{st} \tau_t^{(1)} X_s^T\left(y_t - \sum_{j=1, j\neq s}^p \Gamma_{jt}^{(1)}M_{jt}X_j\right)$ \;  
      \Comment{$\boldsymbol{\text{\bf E}_q\left( \beta_{st} \mid \gamma_{st}=1, y\right)}$}
      \State $\Gamma_{st}^{(1)} \gets \left[ 1 + \left(\Sigma_{st}\right)^{-1/2} \right.$
      \State \hspace{1cm}$\left. \times \exp\left\{ \log(1-\omega_s)^{(1)} -  \log(\omega_s)^{(1)} - \frac{1}{2}  \log(\tau_t)^{(1)}- \frac{1}{2}  \log(\sigma^{-2})^{(1)} -\frac{1}{2}\left(M_{st}\right)^2 \left(\Sigma_{st}\right)^{-1} \right\}\right]^{-1}$ \\
         \State \Comment{$\boldsymbol{\text{\bf E}_q\left( \gamma_{st} \mid y\right)}$}
   \EndFor
    \EndFor
    \State $\omega^{(1)} \gets a^* \oslash (a^*+b^*)$, \quad where \; $a^*= a + \Gamma^{(1)}\mathds{1}_d,$\; $b^*=b - \Gamma^{(1)}\mathds{1}_d + d \mathds{1}_p$\Comment{$\boldsymbol{\text{\bf E}_q\left( \omega^{(1)}_s \mid y\right)}$}
         \State $\mathcal{L}^{\text{old}}(q) \gets \mathcal{L}(q)$,\; $\text{it} \gets \text{it} + 1$
         \State Compute $\mathcal{L}(q)$ (see Appendix \ref{AppLB}) based on the current parameter updates\\
         \Until {$|\mathcal{L}(q) - \mathcal{L}^{\text{old}}(q)| < \text{tol}$ or $\text{it}=\text{maxit}$}
  \end{algorithmic}
  }
\end{algorithm}

\noindent The symbols $\odot$ and $\oslash$ are the Hadamard operators standing for element-wise multiplication and division of two matrices of the same dimension.

\newpage 

\subsection{Computational details}

The convergence characteristics of our procedure can be described in terms of those of any deterministic iterative algorithm. In our experiments, we set the tolerance for the stopping criterion to $10^{-6}$, as our empirical tests suggest that when using a smaller tolerance, the additional time required until convergence does not yield noticeably better inferences. While MCMC sampling may require thousands of iterations to converge, our algorithm usually converges in tens of iterations. As suggested by the runtime profiling provided in Appendix \ref{AppProf}, inference for typical genome-wide association problems with multiple outcomes is usually completed in hours. 
Our algorithm requires the initialization of the variational parameters, but unfortunately comes with no guarantee that it will attain a global minimum for the Kullback--Leibler divergence, $\text{KL}\left(q \middle\| p\right)$. This drawback can be alleviated by using several different initializations, at the price of increasing the computational effort. In practice we did not encounter situations where different starting points gave different optima. The source code can be found in the publicly available R package \texttt{locus}.

\section{Details on the empirical quality assessment of the variational approximation}\label{AppC}

\subsection{Marginal likelihood computation}

We have
\begin{eqnarray}
p(y) &=&  \int\cdots\int \mathrm{d}\omega\, \mathrm{d}\sigma^{-2} \, p( \omega)\, p\left(\sigma^{-2}\right)\nonumber\\ &&\hspace{2.5cm} \times \,\prod_{t=1}^d \left\{\sum_{\gamma_t \in \{0,1\}^p} p\left(\gamma_t \mid\omega\right) \int\cdots\int  \mathrm{d}\beta_t\, \mathrm{d}\tau_t\,   p\left(y_t \mid \beta_t, \tau_t\right) \,p\left(\beta_t \mid \gamma_t, \tau_t, \sigma^{-2} \right)  \,p(\tau_t) \right\} \nonumber\\
&=& \int\cdots\int \mathrm{d}\omega\,\mathrm{d}\sigma^{-2}  \, \left\{ \prod_{s=1}^p\,p( \omega_s)\right\} \,p\left(\sigma^{-2}\right)\prod_{t=1}^d \left\{\sum_{\gamma_t \in \{0,1\}^p}  p\left(y_t\mid\gamma_t, \sigma^{-2}\right)  \prod_{s=1}^p\,p\left(\gamma_{st} \mid\omega_s\right) \right\}\,,
\end{eqnarray}
and one can obtain a closed form expression for $p\left(y_t\mid\gamma_t, \sigma^{-2}\right)$, after integrating out $\beta_t$ and then $\tau_t$. Indeed, proceeding similarly as in \citet{george1997approaches},
\begin{eqnarray} p\left(y_t\mid\tau_t, \gamma_t, \sigma^{-2}\right) &=&
 \int  \left(2\pi\right)^{-n/2} \tau_t^{n/2} \exp\left\{-\frac{\tau_t}{2}\Vert y_t- X\beta_t\Vert^2\right\} \left(2\pi\right)^{-q_{\gamma_t}/2} \sigma^{-q_{\gamma_t}}\tau_t^{q_{\gamma_t}/2}\exp\left\{-\frac{\tau_t}{2} \sigma^{-2} \Vert \beta_t\Vert^2\right\}\mathrm{d}\beta_t\nonumber\\
 &=& \int \left(2\pi\right)^{-n/2-q_{\gamma_t}/2} \tau_t^{n/2+q_{\gamma_t}/2} \exp\left\{-\frac{\tau_t}{2} \left(\beta_{\gamma_t} - \mu_{\beta_t}\right)^T V_{\gamma_t,\sigma^{-2}} \left(\beta_{\gamma_t} - \mu_{\beta_t}\right)\right\} \nonumber\\\nonumber&&\hspace{0.3cm}\times \exp\left( -\frac{\tau_t}{2}S^2_{\gamma_t}\right) \sigma^{-q_{\gamma_t}} \mathrm{d}\beta_t\,,\nonumber
\end{eqnarray}
where 
\begin{eqnarray}
q_{\gamma_t} &=& \sum_{s=1}^p \gamma_{st},\qquad
\tilde{X}_{\gamma_t} = \left( \begin{array}{c} X_{\gamma_t} \\ \sigma^{-1} I_{q_{\gamma_t}} \end{array} \right), \qquad \tilde{y}_t = \left( \begin{array}{c} y_t \\ 0 \end{array}\right)\,,\nonumber\\
S^2_{\gamma_t,\sigma^{-2}} &=& \Vert \tilde{y}_t\Vert^2 - \tilde{y}_t^T \tilde{X}_{\gamma_{t}}  \left(\tilde{X}_{\gamma_t}^T \tilde{X}_{\gamma_t}\right)^{-1} \tilde{X}_{\gamma_{t}}^T \tilde{y}_t = \Vert y_t\Vert^2 - y_t^T X_{\gamma_{t}} V_{\gamma_t,\sigma^{-2}}^{-1} X_{\gamma_{t}}^T y_t\,,\nonumber\\ V_{\gamma_t,\sigma^{-2}} &=& \tilde{X}_{\gamma_t}^T \tilde{X}_{\gamma_t} = X_{\gamma_t}^TX_{\gamma_t} + \sigma^{-2}I_{q_{\gamma_t}}, \qquad
\mu_{\beta_t}=V_{\gamma_t,\sigma^{-2}} ^{-1}\tilde{X}_{\gamma_{t}}^T\tilde{y}_t\,.
\end{eqnarray}
Hence, if $q_{\gamma_t} \neq 0$,
$$p\left(y_t\mid\tau_t, \gamma_t, \sigma^{-2}\right) = \left(2\pi\right)^{-n/2} \tau_t^{n/2} \det\left(V_{\gamma_t,\sigma^{-2}}\right)^{-1/2}\exp\left( -\frac{\tau_t}{2}S^2_{\gamma_t}\right) \sigma^{-q_{\gamma_t}}\,.$$
Now,
\begin{eqnarray} p\left(y_t\mid \gamma_t, \sigma^{-2}\right) &=& \int p\left(y_t\mid\tau_t, \gamma_t, \sigma^{-2}\right) p(\tau_t) \mathrm{d}\tau_t \nonumber\\
&=& \int  \left(2\pi\right)^{-n/2} \tau_t^{n/2} \det\left(V_{\gamma_t,\sigma^{-2}}\right)^{-1/2}\exp\left( -\frac{\tau_t}{2}S^2_{\gamma_t}\right) \sigma^{-q_{\gamma_t}} \frac{\kappa_t^{\eta_t}}{\Gamma(\eta_t)}\tau_t^{\eta_t-1}\exp\left\{-\kappa_t\tau_t\right\} \mathrm{d}\tau_t \nonumber\\
&=& \left(2\pi\right)^{-n/2} \det\left( V_{\gamma_t, \sigma^{-2}}\right)^{-1/2}  \Gamma\left(\frac{n}{2}+\eta_t \right)\frac{\kappa_t^{\eta_t}}{\Gamma(\eta_t)} \left(\kappa_t + \frac{S^2_{\gamma_t}}{2}\right)^{-n/2-\eta_t}\left(\sigma^{-2}\right)^{q_{\gamma_t}/2} \,.\nonumber
\end{eqnarray}
If $q_{\gamma_t}=0$, then

$$p\left(y_t\mid \gamma_t, \sigma^{-2}\right) =  \left(2\pi\right)^{-n/2} \Gamma\left(\frac{n}{2}+\eta_t \right)\frac{\kappa_t^{\eta_t}}{\Gamma(\eta_t)} \left(\kappa_t + \frac{\Vert y_t \Vert^2}{2}\right)^{-n/2-\eta_t}\,.$$

\subsection{Simple Monte Carlo posterior quantities}\label{AppSMCp}

The marginal posterior probability of inclusion for covariate $X_s$ and response $y_t$ can be approximated using simple Monte Carlo sums, as follows,
\begin{eqnarray}
p( \gamma_{st}=1 \mid y ) &=& \frac{p(\gamma_{st}=1, y)}{p(y)} \nonumber\\
&=& \frac{1}{p(y)} \int\cdots\int \mathrm{d}\omega \,\mathrm{d}\sigma^{-2}\, \left\{ \prod_{s=1}^p\,p( \omega_s)\right\}\, p\left(\sigma^{-2}\right) \times \nonumber\\ &&\left[ \prod_{t'\neq t} \left\{\sum_{\gamma_{t'} \in \{0,1\}^p}  p\left(y_{t'}\mid\gamma_{t'}, \sigma^{-2}\right)  \prod_{s'=1}^p\,p\left(\gamma_{s't'} \mid\omega_{s'}\right) \right\}\times \right.
\nonumber\\
&&\hspace{0.2cm}\left. \left\{\sum_{\gamma_{t} \in \{0,1\}^p:\; \gamma_{st}=1}  p\left(y_{t}\mid\gamma_{t}, \sigma^{-2}\right)  \prod_{s'=1}^p\,p\left(\gamma_{s't} \mid\omega_{s'}\right) \right\}\right]  \nonumber\\
&=& \frac{1}{p(y)} \;\frac{1}{I}\sum_{i=1}^I \left[ \prod_{t'\neq t} \left\{\sum_{\gamma_{t'} \in \{0,1\}^p}  p\left(y_{t'}\mid\gamma_{t'}, \left(\sigma^{-2}\right)^{(i)}\right)  \prod_{s'=1}^p\,p\left(\gamma_{s't'} \mid\omega_{s'}^{(i)}\right) \right\}\times \right.
\nonumber\\
&&\hspace{0.2cm}\left. \left\{\sum_{\gamma_{t} \in \{0,1\}^p:\; \gamma_{st}=1}  p\left(y_{t}\mid\gamma_{t}, \left(\sigma^{-2}\right)^{(i)}\right)  \prod_{s'=1}^p\,p\left(\gamma_{s't} \mid\omega_{s'}^{(i)}\right) \right\}\right]\,,\nonumber
\end{eqnarray}
where the samples are generated independently from 
\begin{equation}\left(\sigma^{-2}\right)^{(i)} \sim \text{Gamma}(\lambda, \nu)\,, \qquad \omega_s^{(i)} \sim \text{Beta}(a_s, b_s),\quad s=1, \ldots, p,\qquad i=1, \ldots, I\,.\end{equation}
Similarly, we approximate the posterior mean for $\omega_s$ as
\begin{eqnarray}
\text{E}( \omega_{s} \mid y ) &=& \int \omega_{s}\,  p( \omega_{s} \mid y ) \mathrm{d}\omega_{s} = \frac{1}{p(y)} \int \omega_{s}\, p(\omega_{s}, y) \mathrm{d}\omega_{s} \nonumber\\
&=& \frac{1}{p(y)} \int\cdots\int \mathrm{d}\omega \;\mathrm{d}\sigma^{-2}\;\omega_{s}\; \left\{ \prod_{s'=1}^p\,p( \omega_s')\right\}\,p\left(\sigma^{-2}\right) \times \nonumber\\ &&\prod_{t=1}^d \left\{\sum_{\gamma_t \in \{0,1\}^p}  p\left(y_t\mid\gamma_t, \sigma^{-2}\right)  \prod_{s'=1}^p\,p\left(\gamma_{s't} \mid\omega_{s'}\right) \right\}
\nonumber\\
&=& \frac{1}{p(y)} \;\frac{1}{I}\sum_{i=1}^I \;\omega_{s}^{(i)}\;\prod_{t=1}^d \left\{\sum_{\gamma_t \in \{0,1\}^p}  p\left(y_t\mid\gamma_t, \left(\sigma^{-2}\right)^{(i)}\right)  \prod_{s'=1}^p\,p\left(\gamma_{s't} \mid\omega_{s'}^{(i)}\right) \right\}\,.\nonumber
\end{eqnarray}

\begin{figure}[H]
    \centering
    \noindent\includegraphics[scale=1]{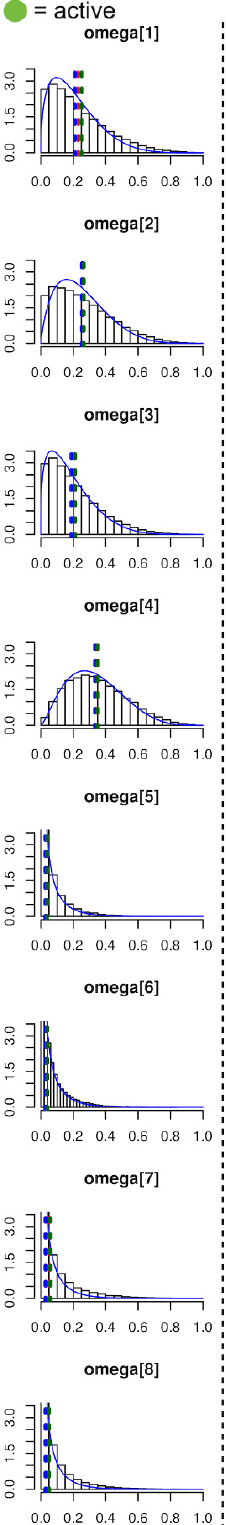}\; 
    \includegraphics[scale=1]{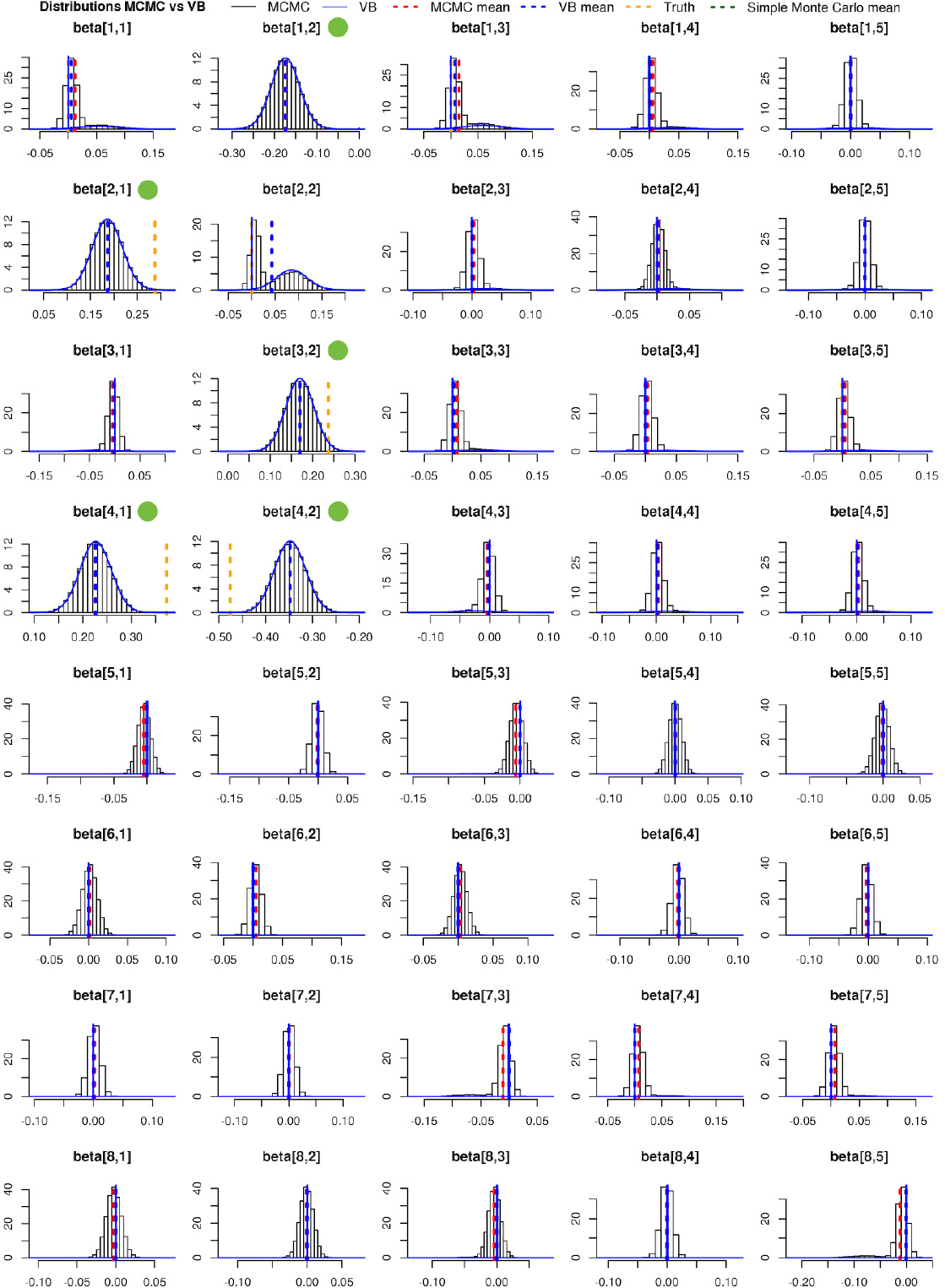}\\ 
    \includegraphics[scale=1]{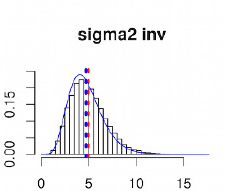}\; 
    \includegraphics[scale=1]{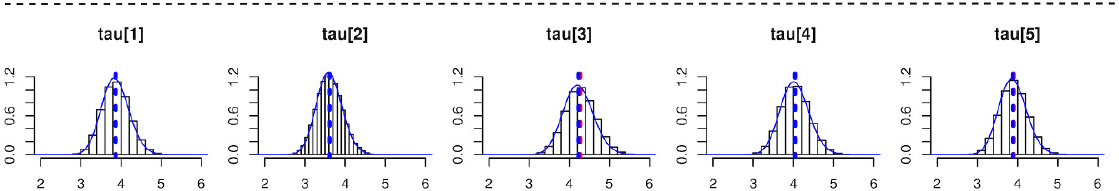} 
    \caption{\small MCMC histograms and variational Bayes (VB, blue) posterior densities for parameters $\omega$ (far left panels), $\beta$ (central panels), $\sigma^{-2}$ (bottom far left panel) and $\tau$ (bottom panels). The MCMC means (dashed red) and variational means (dashed blue) are displayed, along with the simulated values for the $\beta$ plots (dashed orange) and the simple Monte Carlo approximation of the posterior mean of $\omega$ (dashed green). Most of the dashed vertical lines overlap. The problem has of $p=8$ independent covariates and $d=5$ responses for $n=250$ samples. The five green dots indicate the simulated nonzero associations; each explains on average $13.5\%$ of response variance. We use the software OpenBUGS \citep{spiegelhalter2007openbugs} and the R package coda \citep{plummer2006coda} for the MCMC inference and convergence diagnostics.}\label{FigPostDens}
\end{figure}

\begin{figure}[H]
 \centering 
 \noindent\includegraphics[scale=0.96]{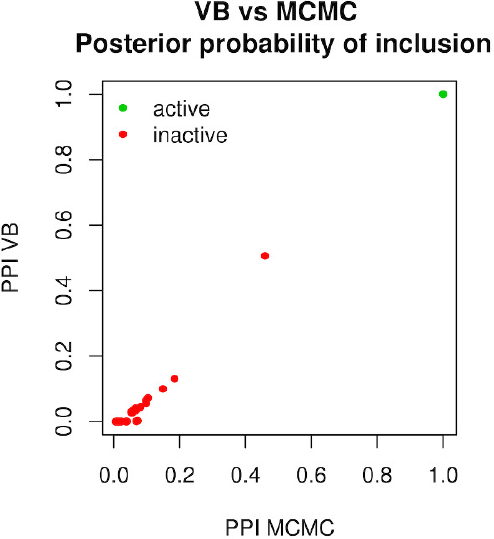} 
  \includegraphics[scale=0.96]{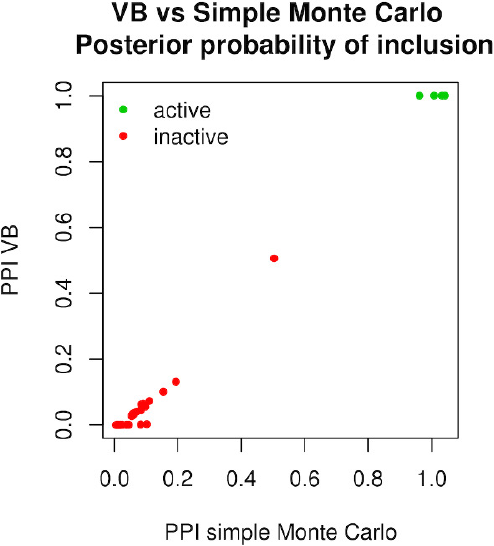}
   \includegraphics[scale=0.96]{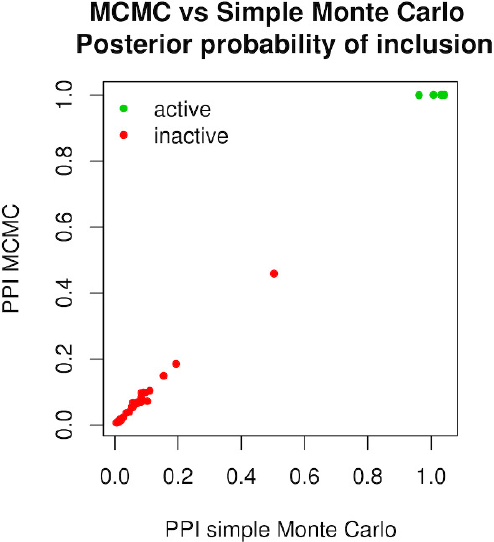}
    \caption{\small Comparison of marginal posterior probabilities of inclusion (PPI) for variational Bayes (VB), MCMC, and simple Monte Carlo approximations. The posterior probabilities of inclusion corresponding to the true signals all overlap, the smallest value being $0.99991$ for variational and $0.99926$ for MCMC inferences. A better simple Monte Carlo approximation might be obtained by increasing the number of draws (here $I=2 \times 10^5$).}\label{FigPPI}
\end{figure}

\begin{figure}[H]
 \centering
\noindent\includegraphics[scale=0.59]{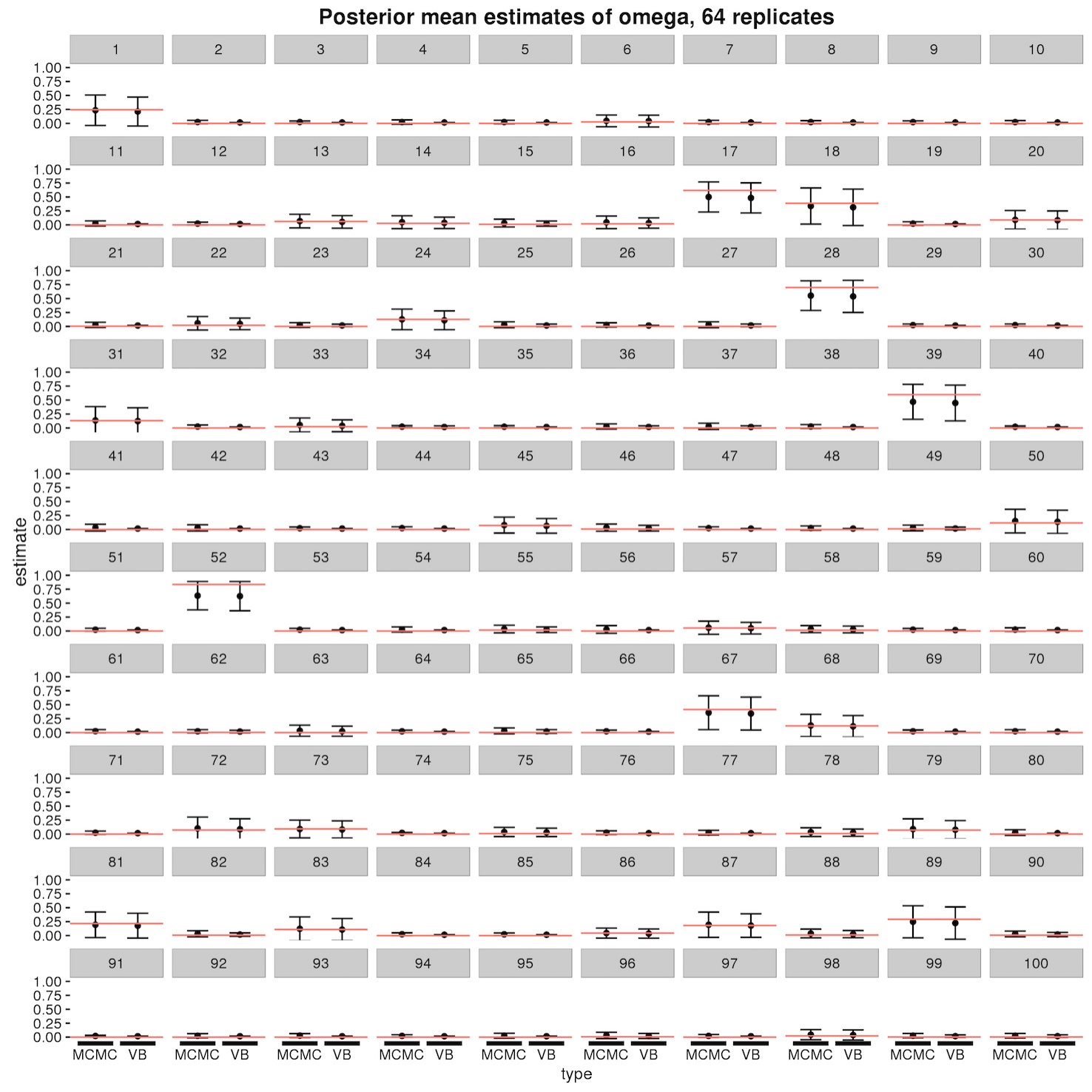}\quad
  \noindent\includegraphics[scale=0.81]{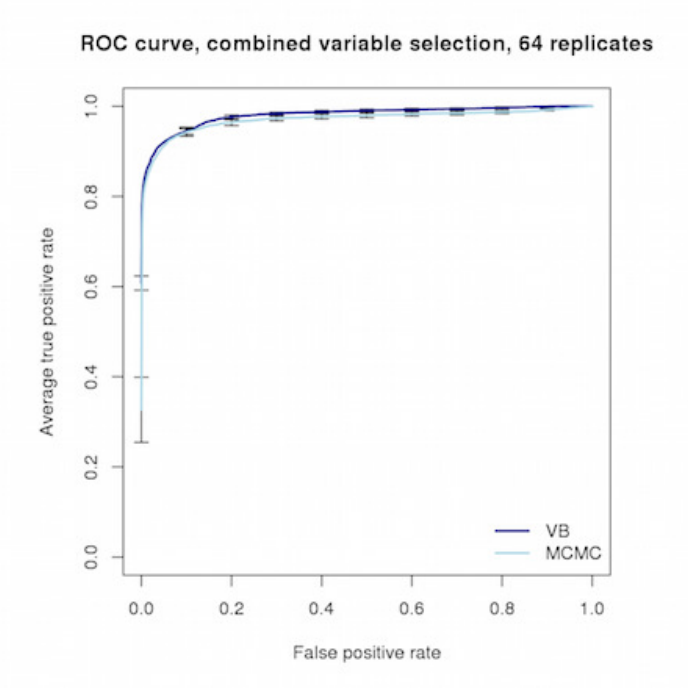}\\ 
\includegraphics[scale=0.59]{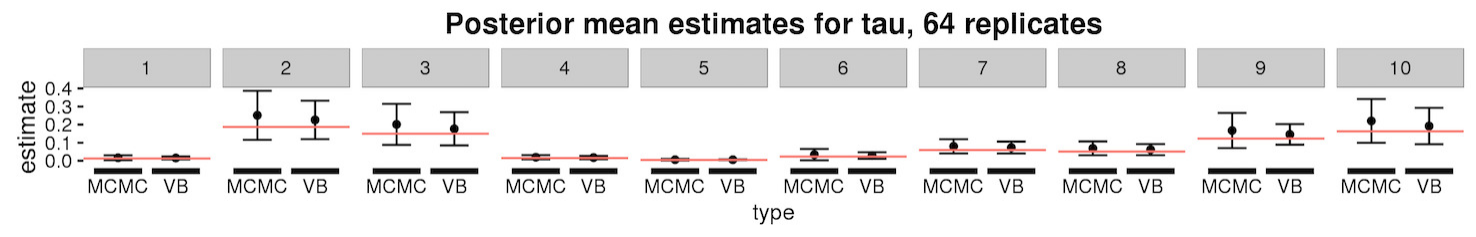}\hspace{5.88cm}
    \caption{\small Comparison of variational and MCMC inferences on data generated from the model with $p=100$ covariates, $d=10$ responses and $n=50$ samples. Left: posterior means of $\omega$ (central panels) and $\tau$ (bottom) with confidence intervals from $64$ replications. The confidence intervals are similar, and most of them contain the simulated value of the parameter (red horizontal lines). Right: average receiver operating characteristic curves for combined variable selection based on the full marginal posterior probabilities matrix, i.e., $\{\gamma^{(1)}_{st}\}$ for variational inference and obtained by dividing the matrix of counts for $\{\gamma_{st}\}$ by the chain length minus the burn-in length for MCMC inference.}\label{FigMod}
\end{figure}

\begin{figure}[H]
 \centering
    \noindent \includegraphics[scale=0.91]{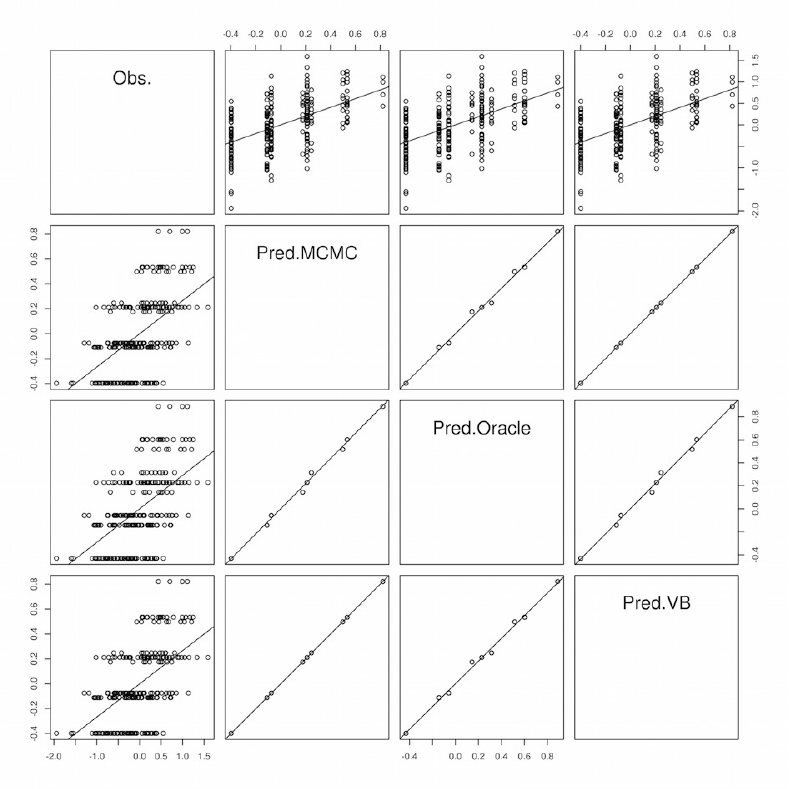} 
  \includegraphics[scale=0.91]{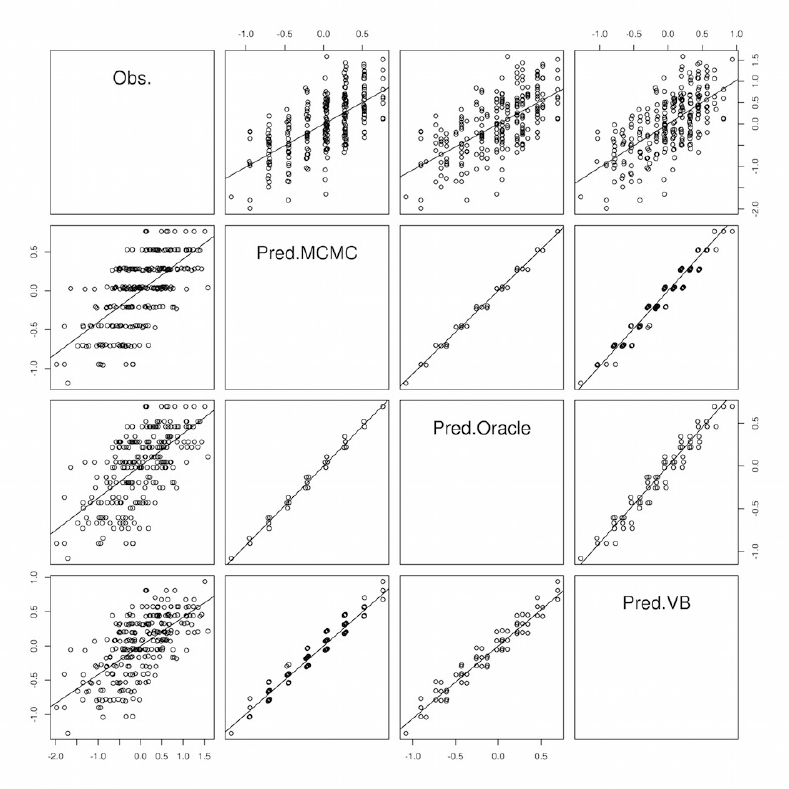}
    \caption{\small Observed values $y_t$ and estimated posterior means of $X \beta_{t}$ obtained by variational and MCMC approximations, and an oracle, for $t=1$ (left panels, two active covariates) and $t=2$ (right panels, three active covariates). Simulation described in Section \ref{SecComp}, $p=8$ covariates, $d=5$ responses and $n=250$ samples.}\label{FigPred}
\end{figure}

Figures \ref{FigPostDens}, \ref{FigPPI}, \ref{FigMod}, and \ref{FigPred} display and compare diverse posterior quantities obtained by variational, MCMC or simple Monte Carlo approximations.

\section{Details on the simulation studies}\label{AppD}

\subsection{Data generation design}\label{AppGen}

We provide here some implementation details on the data-generation settings described in Section \ref{SecPred}. The same general design is used for all the numerical experiments; it is tailored to genetic association studies with multiple outcomes. The dependence structure of simulated SNPs and molecular outcomes is by blocks. As we assume Hardy--Weinberg equilibrium for the SNPs, their marginal distribution is a binomial $\text{B}(2, m)$ with probability $m$ equal to their respective minor allele frequencies, so we model dependence within each block using realisations from a multivariate Gaussian latent variable whose correlation matrix describes a desired dependence structure (autocorrelation with prescribed correlation coefficient or correlation structure corresponding to that of real SNPs at disposal). SNPs are then obtained using a quantile thresholding rule involving their preselected minor allele frequencies; this approach is along the lines of copula-based dependence modelling (without having to resort to copulas). When simulating SNPs that emulate real SNPs data by approximating their minor allele frequencies and correlation structure, the empirical covariance of real SNP blocks may not be positive definite, in which case we approximate it by the closest positive definite covariance matrix (in Frobenius norm) using the algorithm of \citet{higham2002computing} implemented in the R package Matrix \citep{batesmatrix}.
Outcomes are associated with SNPs under an additive dose-effect scheme (uniform and linear increase in risk for each copy of the minor allele) and the proportions of outcome variance explained per active SNP are drawn from a $\text{Beta}(\alpha, \beta)$ distribution, with shape parameters $\alpha=2$ and $\beta=5$ chosen to give more weight to smaller effect sizes. These proportions are rescaled to match a preselected average proportion of outcome variance explained per active SNP (assuming that genetic and external environmental effects are uncorrelated). The magnitude of SNP effects derives from these choices and the sign of these effects is switched with probability $0.5$. Such a design implies an inverse relationship between minor allele frequencies and effect sizes, which is expected to occur since selection against SNPs with large penetrance is stronger \citep[see, e.g.,][]{park2011distribution}. Additional information regarding the generation of a pleiotropic association pattern is given in the paper, as it may vary with the test case considered.

\subsection{Competing predictor selection methods}\label{AppMeths}

The performance of our approach in terms of predictor selection is compared with the following regression procedures: 
\begin{itemize}
\item univariate ordinary least squares: each response $y_t$ is regressed on each covariate $X_s$ and the statistics $\max_{t=1, \ldots, d} t_{st}$, where $t_{st}$ is the $t$-statistic for the significance of $\beta_{st}$, are gathered and ranked;
\item elastic net regression for multivariate Gaussian responses ($\alpha=0.5$) with 10-fold cross-validation for choosing the tuning parameter $\lambda$~\citep[glmnet,][]{friedman2009glmnet}. The estimates $|\beta_{s}|$ ($s=1, \ldots, p$), where $\beta_{s}$ is the regression coefficient estimate for $X_s$ and common to all responses, are gathered and ranked;
\item univariate Bayesian regressions, lmBF \citep{morey2015package}: each response $y_t$ is regressed on each covariate $X_s$ with all computations made analytically. 
The (average) Bayes factors, $\sum_{t=1}^d \text{BF}_{st}/d$ ($s=1, \ldots, p$), are gathered and ranked; 
\item $d$ Bayesian multiple regressions, BAS \citep{clyde2010predict}, one for each response, using MCMC inference. A g-prior is used for the regression coefficients. The (average) Bayes factors, $\sum_{t=1}^d \text{BF}_{st}/d$ ($s=1, \ldots, p$), are gathered and ranked; and
\item $d$ Bayesian multiple regressions, varbvs \citep{carbonetto2012scalable}, one for each response, using variational inference. The posterior probabilities of inclusion are summed across responses and ranked.
\end{itemize}

\subsection{Runtime profiling}\label{AppProf}

\begin{figure}
\centering
 \noindent \includegraphics[scale=0.85]{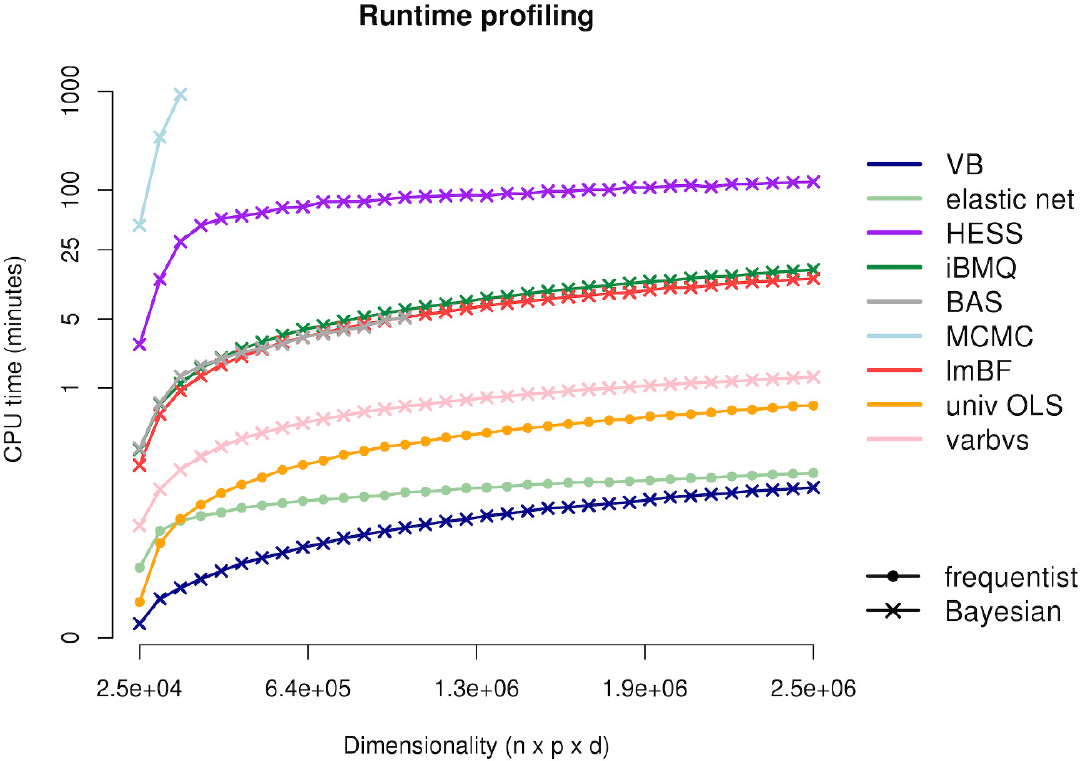} 
    \caption{\small Serial runtime profiling for all methods discussed, on an Intel Xeon CPU at 2.60 GHz with 64 GB RAM.}\label{FigProf}
\end{figure}

We report a runtime profiling of the different methods for a range of problem sizes ($n\times p \times d$). Figure \ref{FigProf} displays the run times in minutes, averaged over $24$ replications. We do not aim to provide precise and exhaustive comparisons, since the methods all depend on parallelism and convergence characteristics that are not directly comparable. All methods were run serially, in an attempt to treat them on an equal footing. This choice could be challenged, as some approaches are more parallelizable than others, but the number of cores used for the latter represents an additional setting that would come into play with a potentially large impact on the measures. The number of chains for MCMC inferences also matters: HESS \citep{richardson2010bayesian} is run with three chains (following its authors' choice made in their simulations); the other MCMC inferences are based on a single chain. Finally, the runtime may also greatly vary depending on the chosen chain length: the latter is adaptively selected by the Bayesian multiple regression method BAS \citep{clyde2010predict}, and, based on preliminary convergence diagnostics, it is set to $50,000$ samples for HESS, iBMQ \citep{scott2012integrated} and the MCMC inference on our model using OpenBUGS \citep{spiegelhalter2007openbugs}. In practice, the number of samples needed until convergence may increase greatly with the problem size, a fact that was not accounted for in this profiling. Hence, by adequately increasing the chain lengths when considering larger dimensionalities, we expect the curves of Figure \ref{FigProf} corresponding to the MCMC approaches to deviate more widely from that of our method. 
With these serial settings, our approach is faster than all Bayesian and frequentist methods. MCMC inference for our model is the slowest of all tested methods, which underlines the intractability of MCMC sampling for large problems. Inference for HESS is faster, but still more than $650$ times slower than our variational approach. Our method is also about $10$ times faster than $d$ applications (one for each outcome) of the varbvs method \citep{carbonetto2012scalable} but the runtime of the varbvs procedure can be reduced using multiple cores.

\section{Details on the real data problem}\label{AppE}

\subsection{Clinical study and preprocessing of the mQTL data analysis}\label{AppPre} 

Diogenes \citep{2010_Diogenes_Obesity_rev} is a large clinical dietary intervention study, which enrolled roughly $1,000$ overweight subjects from $8$ European centers. The subjects were assigned to a $8$-week weight-loss program followed by $6$ months of supervised ad-libitum diet, meant as a weight-maintenance phase. For the $6$-month period, individuals were randomized into five intervention groups, whose diets differed in their macronutrient and glycemic load content. The clinical trial had the broad objective of elucidating whether certain macronutrient compositions are more successful in weight maintenance than others. It gathered data on genetic variants, gene, protein and metabolite expression. 
In addition, more than $7,000$ clinical, anthropometric and behavioural variables were made available for each individual. 
When meaningful to do so, the data were collected at three time points, reflecting different stages of the dietary treatment. 

In Section \ref{SecReal}, we performed an mQTL analysis which involves tag SNPs from Illumina HumanCore chips (about 300k SNPs) and metabolites from plasma obtained by liquid chromatography-mass spectrometry (LC-MS). We excluded SNPs having  missing values, minor allele frequency below $5\%$, call rate below $95\%$, or violating the Hardy--Weinberg equilibrium. Moreover, subjects having gender discrepancy (e.g., subject recorded as male but being homozygous for each X chromosome marker), abnormal autosomal heterozygosity or whose genomes were too close to each other (IBS$>95\%$) were also excluded. Additional subjects were removed after applying the Tukey method for outlier detection \citep{tukey1949comparing}, based on the metabolomic data. Metabolite expression levels were log2-transformed, and had no missing values. After these quality checks, the data consisted of $p=~215'907$ SNPs and $d=125$ metabolite expression levels, for $n=317$ individuals.

\subsection{Permutation-based Bayesian false discovery rate estimation}\label{AppFDR} 

We provide details on the false discovery rate estimation procedure applied in Section \ref{SecReal} to compare our method with the varbvs method of \citet{carbonetto2012scalable} on the real data. We use the two-group mixture approach proposed by \citet{efron2008microarrays} in the context of microarray data analysis: we simultaneously consider $N$ null hypotheses and their corresponding test statistics, which we assume to follow a mixture distribution
$$F = \pi_0 F_0 + (1-\pi_0) F_1\,,$$
where $\pi_0$ is the prior probability for a null case, and $F_0$ and $F_1$ are the null and non-null cumulative distribution functions. The ``Bayesian false discovery rate'' for some threshold $\tau$ is the posterior probability that a rejected null hypothesis (i.e., test statistic exceeding $\tau$) is a false positive,
\begin{equation}\label{SMEqFDRth}\textup{FDR}(\tau) = \frac{\pi_0 \bar{F}_0(\tau)}{\bar{F}(\tau)}\,,\end{equation}
where $\bar{F} = 1- F$ and $\bar{F_0} = 1- F_0$. We derive estimates of (\ref{SMEqFDRth}) based on the posterior probabilities of inclusion quantifying the associations between each covariate-response pair. Specifically, we obtain an empirical null distribution by running our algorithm (with the same hyperparameters as those used for the actual inference) on $B$ datasets with permuted outcome sample labels and compute, for a grid of thresholds $0<\tau_1 < \ldots < \tau_K<1$,
\begin{equation}\label{SMEqFDR}\widehat{\textup{FDR}}(\tau_k) = \frac{\textup{median}_{b = 1, \ldots, B} \#\{\textup{PPI}_{st}^{(b)} > \tau_k; \; s=1, \ldots, p;\; t = 1, \ldots, d \}}{\#\{\textup{PPI}_{st}>\tau_k; \; s=1, \ldots, p;\; t = 1, \ldots, d\}}\,,\qquad k = 1, \ldots, K,\end{equation}
where we conservatively set $\pi_0$ to $1$ in (\ref{SMEqFDRth}). As the posterior probabilities of inclusion obtained by applying varbvs $d$ times (one multiple regression for each outcome) are not identically calibrated across all $d$ estimations, we use adaptive thresholds on the columns of the varbvs posterior probabilities of inclusion matrix, $$\tau_k(y_t) =  \frac{\textup{median}_{s}\left(\textup{PPI}_{st}\right)}{\textup{median}_{s,t'}\left(\textup{PPI}_{st'}\right)}\, \tau_k\qquad t = 1, \ldots, d, \quad k = 1, \ldots, K.$$ 
To find thresholds $\hat{\tau}$ corresponding to preselected false discovery rates, we fit a cubic spline to the estimates (\ref{SMEqFDR}) previously obtained for $\tau_1, \ldots, \tau_K$.

\subsection{Biological evidence for the mQTL analysis findings}\label{AppBio} 

We support the findings obtained by our method for the mQTL data analysis with external association results from the following online databases:
\begin{itemize}
\item GWAS catalog \citep{welter2014nhgri};
\item UCSC genome browser \citep{karolchik2003ucsc};
\item GTEx \citep{lonsdale2013genotype}; and
\item GeneCards \citep{rebhan1998genecards}.
\end{itemize}
We find direct or indirect links with metabolic activities for $12$ of the $25$ SNPs declared as ``active' by our method: SNPs $rs3820711$, $rs4316911$, $rs4909818$, $rs4744227$, $rs174535$, $rs680379$, $rs8012466$, $rs4906771$, $rs573922$, $rs3903703$, $rs8114788$ and $rs6001093$ have been identified as associated with BMI, diverse diabetic or obesity diseases, fatty acid, sphingolipid or phospholipid levels, either from direct genome-wide association analyses or through protein coding genes for which they were reported as eQTLs.

\subsection{Replication of the mQTL data analysis}\label{AppRepl}

In this appendix, we replicate the real data analysis of Section \ref{SecReal} on a simulated dataset (with twice as many outcomes and slightly more observations), which can be found on Figshare (\url{https://dx.doi.org/10.6084/m9.figshare.4509755.v1}). The analysis can be reproduced using the code available on GitHub (\url{https://github.com/hruffieux/mQTL_analysis_example}), provided that a machine with adequate RAM memory (about $400$G for this large problem) is used.

To best mimic the real mQTL data used in Section \ref{SecReal}, we simulate $p = 215,907$ SNPs based on the sample minor allele frequencies of the real tag SNPs and we reproduce their dependence structure by blocks of $1,000$ consecutive SNPs according to the discussion of Appendix \ref{AppGen}. We also simulate $d=250$ normally distributed outcomes with equicorrelation by blocks using correlation coefficient $\rho \in \{0.5, 0.6, 0.7\}$; this block structure is similar to that of the real metabolite data. To induce a realistic pleiotropic pattern, we simulate associations between $750$ SNPs and $175$ metabolites (randomly chosen) taking the block-wise dependence structure of the latter into account: the probability that a given active SNP is associated with a given active outcome varies across blocks, so correlated metabolites are either all highly likely or all less likely to be associated with the SNP. The average proportion of metabolic variance explained with each association is $2.5\%$ (but more associations explain less than this, as discussed in Appendix \ref{AppGen}). We generate $n=350$ observations.

\begin{figure}
\centering
  \includegraphics[scale=0.66]{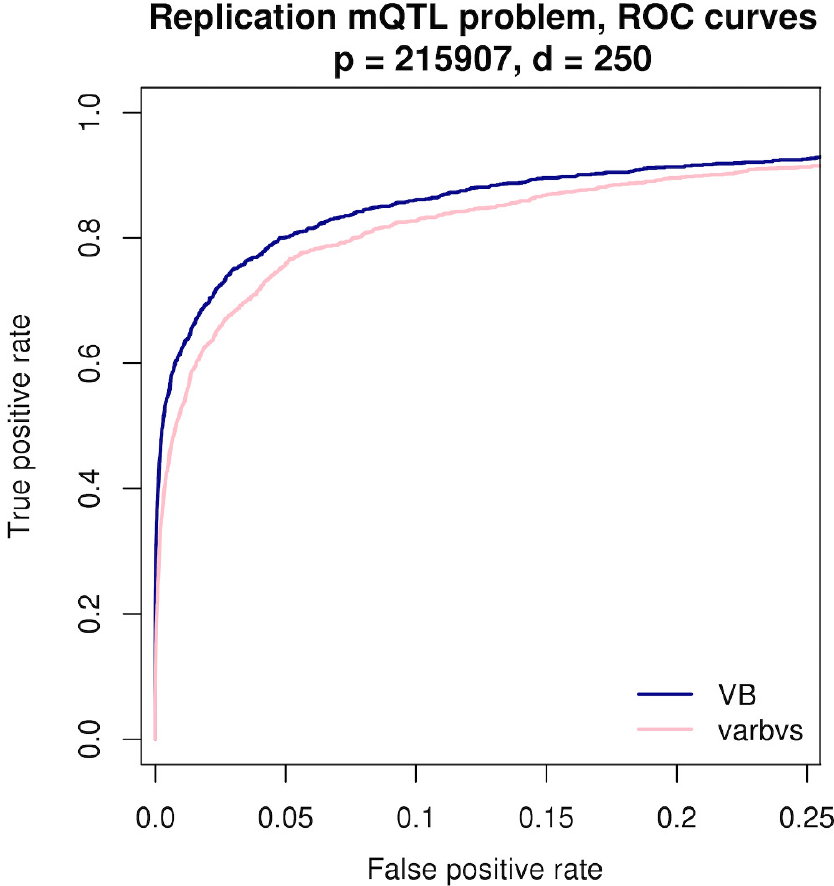}
    \caption{\small Receiver operating characteristic curves based on posterior probabilities of inclusion inferred by VB and by varbvs for simulated mQTL data.}\label{SMFigROCRepl}
\end{figure} 

\begin{table}
\begin{center}
\captionof{table}{\small Replication of the mQTL data analysis. Number of true positives (TP) detected by VB and by varbvs, and number of TP in common at selected permutation-based false discovery rates. \label{SMTabFDR}}
\footnotesize
\begin{tabular}{cccc}
\hline
&\# TP: \\
Permutation-based FDR (\%) & VB & varbvs & VB $\cap$ varbvs \\ 
  \hline
5 & 13 & 18 & 13  \\ 
  10 & 26 & 26 & 23 \\ 
  15 & 32 & 32 & 28 \\ 
  20 & 38 & 32 & 28 \\ 
  25 & 47 & 33 & 29 \\ 
   \hline
\end{tabular}
\end{center}
\end{table}

\begin{figure}
\centering
  \includegraphics[scale=0.73]{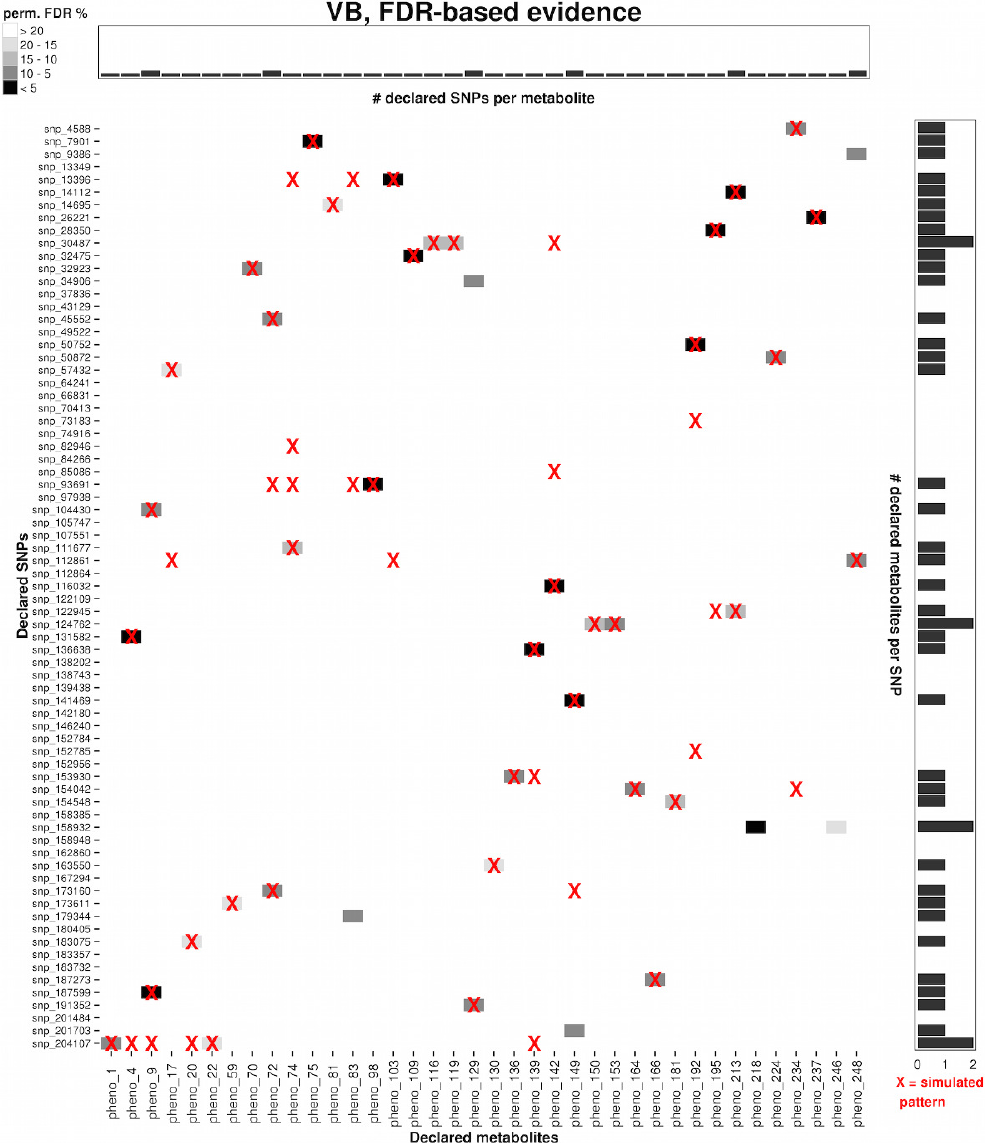}\;\;
\includegraphics[scale=0.73]{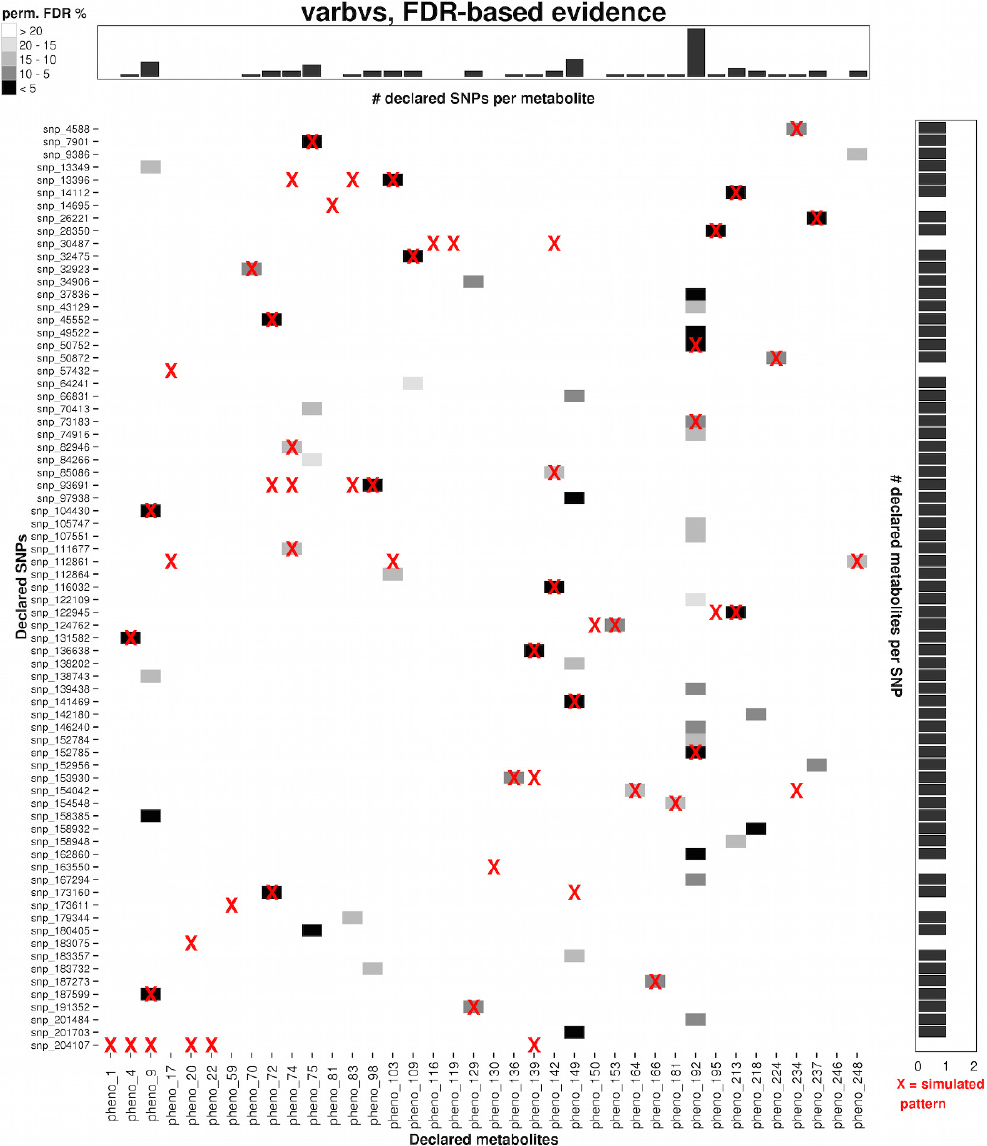}
    \caption{\small Comparison of the associations declared by VB (left) and varbvs (right) at FDR of $20\%$ estimated using $B=200$ permutations for simulated mQTL data; see also Table~\ref{SMTabFDR}. The simulated association pattern is overlaid (red crosses). The large number of false positives on the right plot indicates that the permutation-based FDR estimates are somewhat underestimated for varbvs; an improved estimation based on more permutations would further emphasize the improvement of our method over varbvs.}\label{SMFigCompRepl}
\end{figure} 

\begin{figure}
\centering
\includegraphics[scale=0.81]{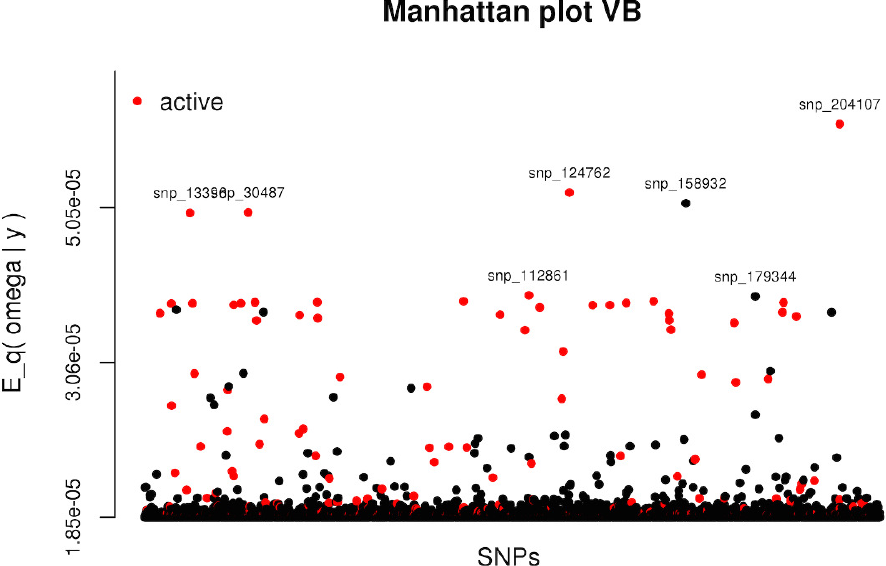}\quad
  \includegraphics[scale=0.81]{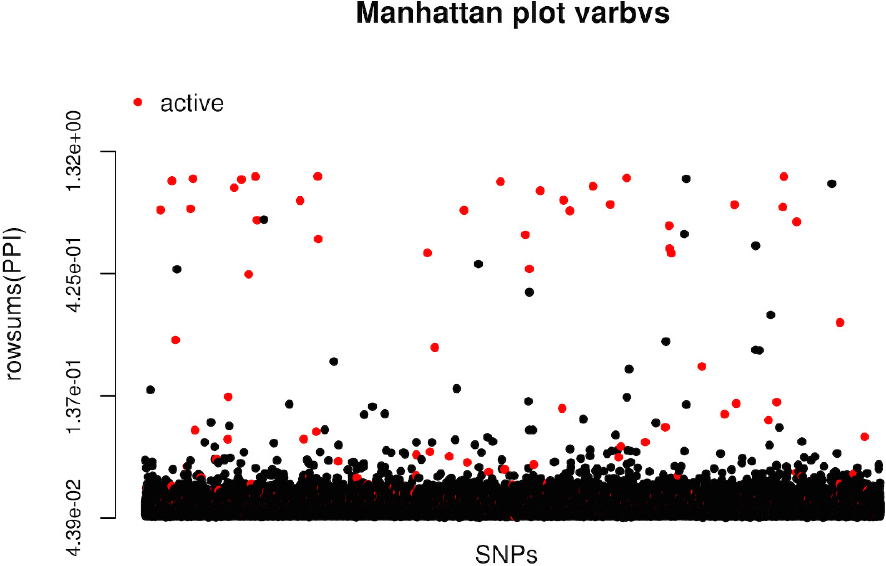}
    \caption{\small Manhattan plots of SNP association and evidence of pleiotropy obtained with VB (left) and varbvs (right) for simulated mQTL data.}\label{SMFigManRepl}
\end{figure} 

We apply the permutation analysis described in Section \ref{SecReal} to varbvs and our method and again obtain a more powerful selection for our method compared to varbvs at estimated FDR of $20\%$ and $25\%$ (Table \ref{SMTabFDR} and Figure \ref{SMFigCompRepl}). Because simulated data are used, we can further support the overall superiority of our method with ROC curves, see Figure \ref{SMFigROCRepl}.

\end{document}